\documentclass{mn2e}

\usepackage[dvips]{graphicx}
\usepackage[fleqn]{amsmath}
\newcommand{\be}{\begin{equation}}
\newcommand{\ee}{\end{equation}}

\title[Wavelet analysis of Cygnus X-1]
      {Wavelet analysis of millisecond variability of Cygnus X-1 during 
       its failed state transition}

\author
 [Lachowicz \& Czerny]
 {\noindent
 Pawe\l\ Lachowicz$^1$\thanks{E-mail: paulo@camk.edu.pl} and
 Bo\.zena Czerny$^1$ \\
 $^1$Nicolaus Copernicus Astronomical Center, Bartycka 18, 00-716 Warszawa, Poland \\
 }

\pagerange{\pageref{firstpage}--\pageref{lastpage}}
\pubyear{2005}

\begin{document}

\maketitle

\label{firstpage}


\begin{abstract}
Application of multi-resolution analysis adopting wavelets in order to investigate 
millisecond aperiodic X-ray variability of Cyg X-1 is presented. This relatively new 
approach in a time-series analysis allows us to study both the significance of any 
flux oscillations localized in time--frequency space, as well as the stability of its 
period. Using the observations of the {\it RXTE}/PCA we analyze archival data from Cyg X-1 
during its failed state transition. The power spectrum of such a state is strongly 
dominated by a single Lorentzian peak corresponding to a damped oscillator. Our wavelet 
analysis presented in this paper shows the existence of short lasting oscillations without 
an obvious trend in the frequency domain. Based on this result, we suggest the interpretation
of the dominant components of Cyg~X-1 power spectrum in all spectral states: 
{\it (i)} a power law component dominating a soft state is due to accretion rate fluctuations 
related to MRI instability in the cold disk, 
{\it (ii)} the low frequency Lorentzian is related to the same fluctuations but in the inner ion 
torus, and 
{\it (iii)} the high frequency Lorentzian (dominating the power spectrum during the failed state 
transition) is due to the dynamical pulsations of the inner ion torus.

The usage of the wavelet analysis as a potential and attractive tool in order to spot the 
{\it semi-direct} evidences of accretion process onto black-holes is proposed.
\end{abstract}


\begin{keywords}
accretion, accretion disks -- binaries: general -- stars: individual: Cyg~X--1
-- X-rays: observations -- X-rays: stars.
\end{keywords}


\section{Introduction}
\label{s:s1}

Cyg X-1 (Bowyer et al.\ 1965) belongs to the class of Galactic accreting black-hole binaries
characterized by a highly variable X-ray emission in time-scales covering up to ten orders of 
magnitude (e.g. Nowak et al.\ 1999; Reig, Papadakis \& Kylafis 2002; Psaltis 2004). 
The optical companion, HDE 226868 star (Bolton 1972; Webster \& Murdin 1972) was classified 
as O9.7Iab supergiant (Walborn 1973) with the orbital period of 5.6 d. The main source of the
matter accreting onto black hole is the focused stellar wind taking its origin from HDE 
226868. Mass ranges of the components are 10--32 M$_\odot$ and 16--55 M$_\odot$ for 
the black-hole and the supergiant, respectively (Gies \& Bolton 1986; Herrero et al.\ 1995; 
Zi\'o{\l}kowski 2005).

The power spectrum density (PSD) of the X-ray emission is broad-band and aperiodic. It 
was frequently characterized as a broken power-law (Sutherland, Weiskopf \& Kahn 1978;
Belloni \& Hasinger 1990; Nowak et al. 1999; Revnivtsev, Gilfanov \& Churazov 2000), 
with occasional narrow quasi-periodic features (QPO) (Angelini, White \& Stella 1992). 
Higher quality data allowed for more sophisticated parameterization of the PSD through a 
number of broad Lorentzians and a power-law component (e.g. Nowak 2000; Pottschmidt et al. 
2003, hereafter P03). The lightcurve is not stationary. In short time-scales the variance is 
proportional to the mean luminosity of the source (Uttley \& McHardy 2001; Gleissner et al. 
2004) so the normalized power spectrum is broadly used to characterize the source (Miyamoto et 
al. 1992 and most of the subsequent papers). In long time-scales also the shape of the PSD changes as 
it depends significantly on the spectral state of the source (e.g. Cui et al. 1997a; 
1997b). 

Although frequently used, Fourier-based technique is actually not suitable for the  
analysis of the aperiodic variability. The Fourier analysis uses a fixed sinusoidal basis 
functions to decompose a signal. This method therefore provides excellent 
frequency resolution of persistent features but it fails when the signal is highly 
aperiodic and existing modulations are needed to be localized in time--frequency space.

To overcome this problem one needs to use a multi-resolution approach, e.g. by wavelet 
analysis (WA). Because the wavelet decomposition gives an idea of both the local frequency 
content of a time-series and the temporal distribution of these frequencies, its 
application in astronomical signal processing is increasing rapidly (e.g. Szatmary, 
Vinko \& Gal 1994; Escalera \& MacGillivray 1995; Frick et al. 1997; Aschwanden et al. 
1998; Barreiro \& Hobson 2001; Freeman et al. 2002; Irastorza et al. 2003).  
Interestingly, so far the application of wavelets in order 
to study variability of X-ray binary systems and active galactic nuclei was marginal. 
Scargle et al. (1993) used them to examine QPO and very low-frequency noise for Sco X-1 
whereas Steiman-Cameron et al. (1997) supplemented Fourier detection of quasi-periodic 
oscillations in the optical lightcurves of GX 339--4 by scalegrams, a wavelet technique. 
Fritz \& Bruch (1998) applied this technique to the analysis of the optical luminosity 
variations in cataclysmic variables while Liszka, Pacholczyk \& Stoeger (2000a) 
analyzed the ROSAT lightcurve of NGC~5548 in short time-scales.

 \begin{figure}
 \begin{center}
 \includegraphics[width=6cm,angle=-90]{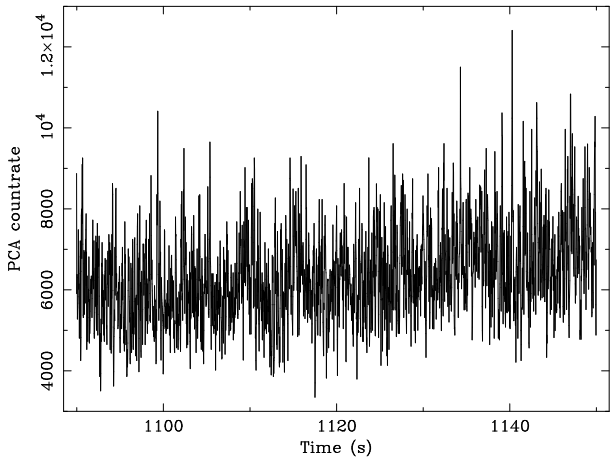}
 \caption
  {
   The lightcurve of Cyg X-1 as observed by the {\it RXTE}/PCA on 1999.12.05, showing
   a 60 s sequence of data selected by us for deeper studies using wavelet analysis.
  }
 \label{fig:krzywa}
 \end{center}
 \end{figure}

In this paper we select a very special observation of Cyg~X-1 performed on 1999.12.05. 
It was examined by P03 and the state of the source was described as 
failed state transition. The power spectrum at that time was particularly simple since the 
dominant part of the power was contained in a single Lorentzian peak. The aim of our 
paper is to verify with the wavelet analysis whether indeed a simple damped oscillator 
well represents the time-resolved lightcurve properties.

Our paper is organized as follows. Section 2 provides information on our data selection 
and reduction. In Section~3 we describe most essential aspects of the time--frequency 
analysis, starting from a brief discussion of computing X-ray power spectra and 
associated time--resolution problem. Next, we provide a description of calculation of 
a continuous wavelet transform in the frame of our interest. Section~4 contains the analysis 
of Cyg X-1 lightcurve whereas a discussion of results, including the lightcurve 
simulations, is the subject of Section~5. We conclude in Section~6.


\section{Data selection and reduction}
\label{s:s2}

To probe the object variability in time-scales of seconds we use the observation by 
Proportional Counter Array on board of the {\it Rossi X-ray Timing Explorer} ({\it 
RXTE}/PCA; Bradt, Rothschild \& Swank 1993) taken from the public archive of the HEASARC. 
We select the PCA data set of 40099-01-24-01 (1999.12.05) on which we concentrate our 
investigation. The data were reduced with the LHEASOFT package v. 5.3.1 (May 2004) 
applying a standard data selection, i.e. the Earth elevation angle $>10^{\circ}$, 
pointing offset $<0^{\circ}.01$, the time since the peak of the last South Atlantic 
Anomaly SAA $>30$ min and the electron contamination $<0.1$. The number of Proportional 
Counting Units (PCUs) opened during the observation was equal 3.

We extract the lightcurve in 2.03--13.1 keV energy band (channels 0--30) and rebin it 
with one common bin size value of $\Delta t=0.025$~s. The average count-rate in this set 
is $\bar{x}=6061.8$ cts~s$^{-1}$ and the fractional rms 21.6 \%. For needs of computation 
of the Fourier power spectra, we use {\sc powspec} software with an applied Poisson noise level 
subtraction. For purposes of wavelet analysis, the IDL software provided by Torrence \& 
Compo (1998) (hereafter TC98; see Acknowledgments for the URL address) was run.


\section{Time--frequency analysis}
\label{s:s3}

Computation of the Fourier power spectral density is an ideal tool for 
detecting periodic fluctuations in the time-series. Therefore, in the absence of major problems 
caused by a window pattern (aliasing) and at high signal-to-noise ratio, a strong peak in 
the Fourier spectrum has an immediate interpretation (e.g. Gray 1976). 

However, the Fourier analysis fails when time evolution of spectral features has to be taken 
into account. In order to capture this effect one needs to apply a 
time--frequency representation of the signal. The simplest solution to this problem is the 
application of Windowed Fourier Transform (see e.g. Cohen 1995; Flandrin 1999 for
review). More advanced methods include inter alia Wigner-Ville distribution, the reassignment
method or the Gabor spectrogram (see Vio \& Wamsteker 2002 and references therein for a nice
review). In the present paper we apply the alternative approach: the wavelet analysis.

\subsection{Fourier power spectrum}
\label{ss:fps}

For a reference, we calculate the PSD of the studied lightcurves using a standard approach (see van der Klis 1989 for review). We use the normalization of Miyamoto et al. (1992) which gives the periodogram in units of (rms/mean)$^2$/Hz. Therefore, the integrated periodogram yields the square of fractional rms of a lightcurve. We subtract the noise as in Vaughan et al. (2003), omitting the correction for dead-times (see Zhang et al. 1995 for the discussion of the dead-time effect), as our data sets
are rather short (20 s) and thus the uncertainty in the determination of the PSD dominates.

\subsection{Wavelet analysis}
\label{ss:wa}

A basic concept standing for its application to the time-series was to analyze the signal at 
different frequencies with different resolutions (see e.g. Farge 1992; Daubechies 1992 and 
TC98 for good summary on wavelets). Contrary to the Fourier transform, the wavelet 
analysis makes use of a set of functions, {\it wavelets}, which are localized in scale--time 
space\footnote{In the wavelet analysis a term {\it scale} is used instead of {\it frequency} 
which is reserved for the Fourier analysis.}. Several standard shapes are broadly used, 
depending on the subject under study.

\subsubsection{Wavelets}
\label{sect:wavelets}

The basic concept is a mother wavelet function, $\psi(t)$, which depends on time-parameter $t$. A 
wavelet's family is therefore generated by scaling and shifting of $\psi(t)$ as follows:
\be
    \psi_{\rm a,b}(t) = \frac{1}{\sqrt{a}}\ \psi\left(\frac{t-b}{a}\right)
    \ \ \ \ \ a,b \in \Re,\ a>0
\label{psiab}
\ee
where $a$ and $b$ are factors standing for scaling and time-shifting, respectively.

 \begin{figure}
 \begin{center}
 \includegraphics[width=6cm,angle=-90]{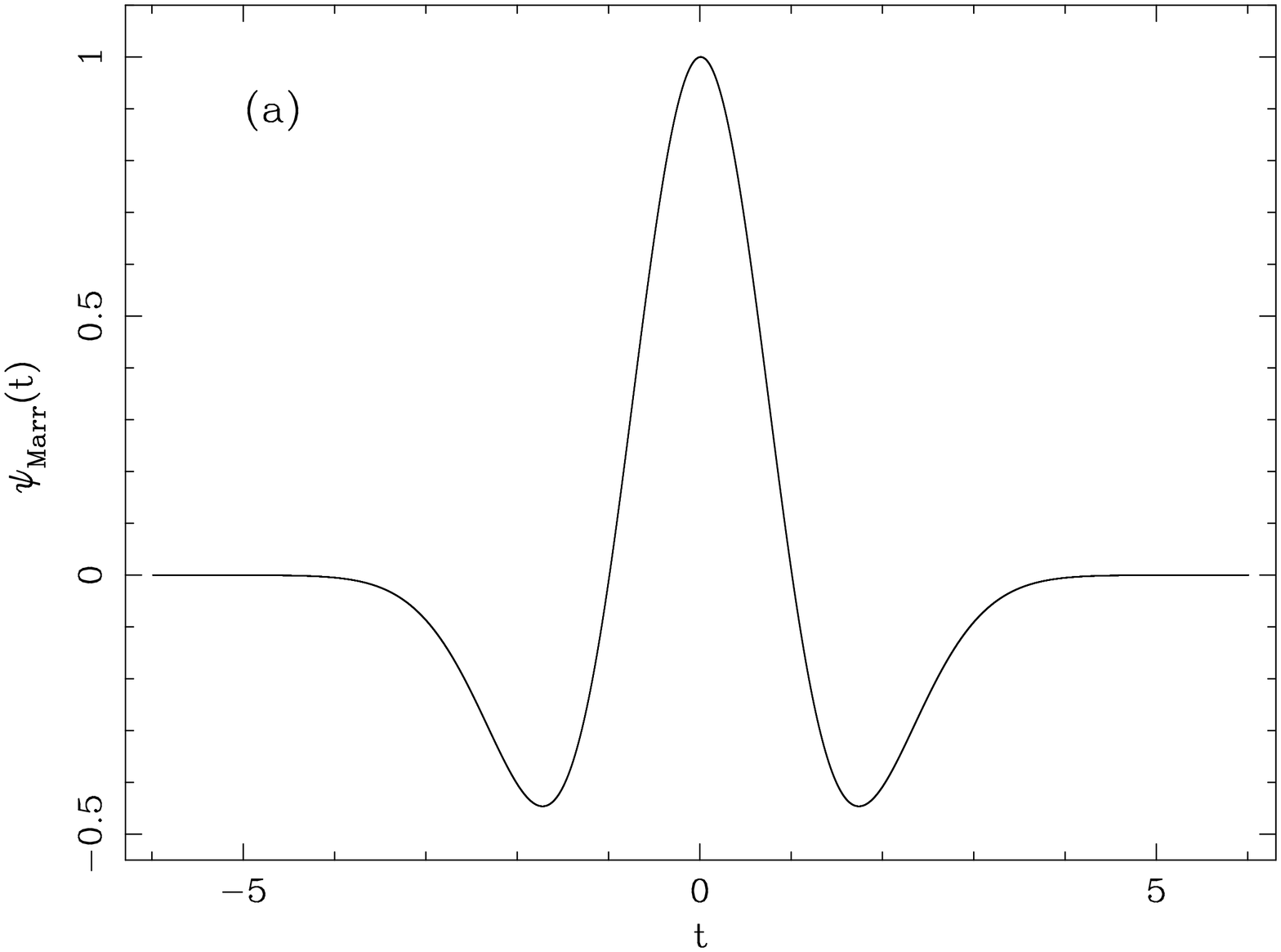}
 \includegraphics[width=6cm,angle=-90]{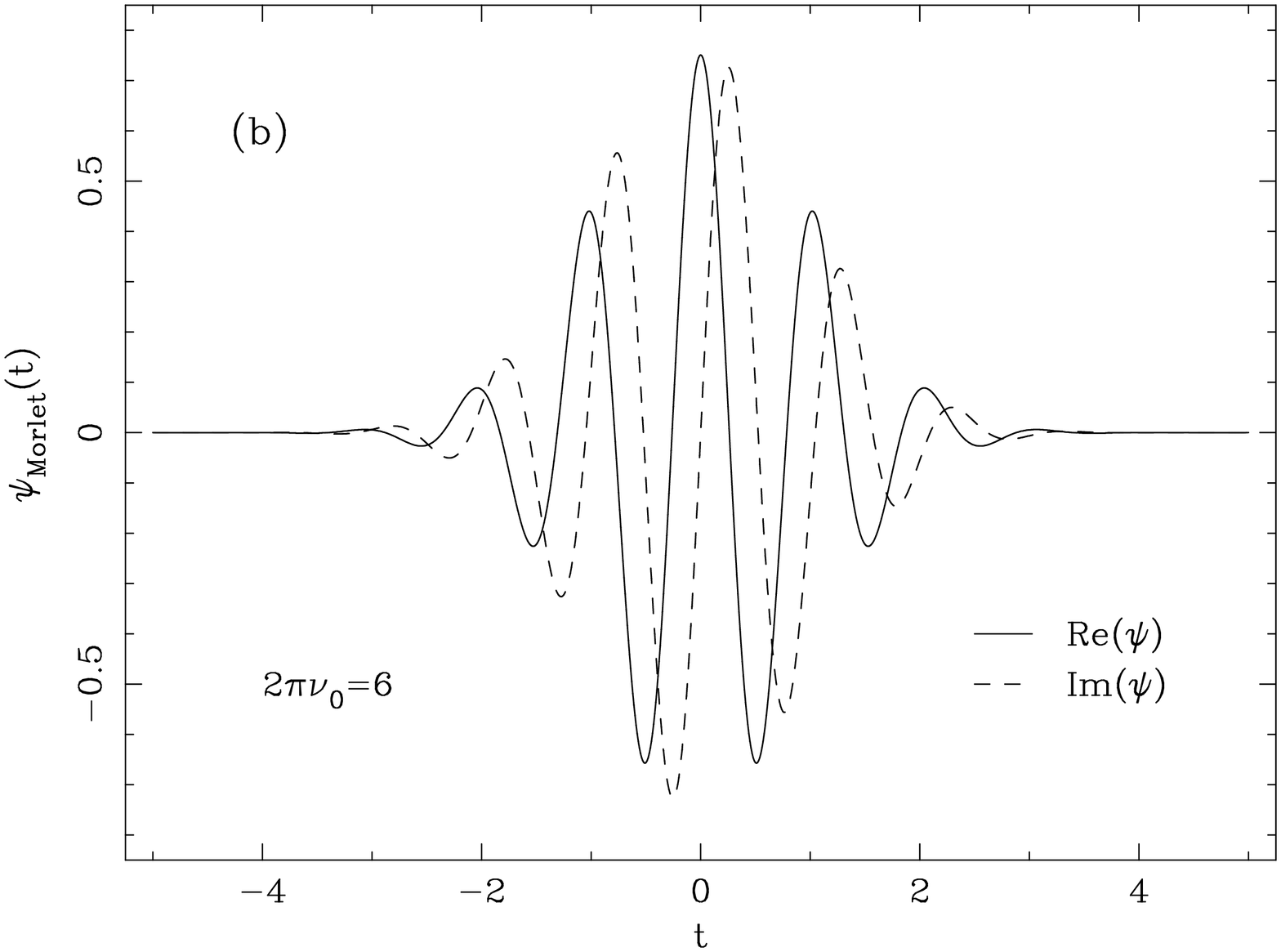}
 \includegraphics[width=6cm,angle=-90]{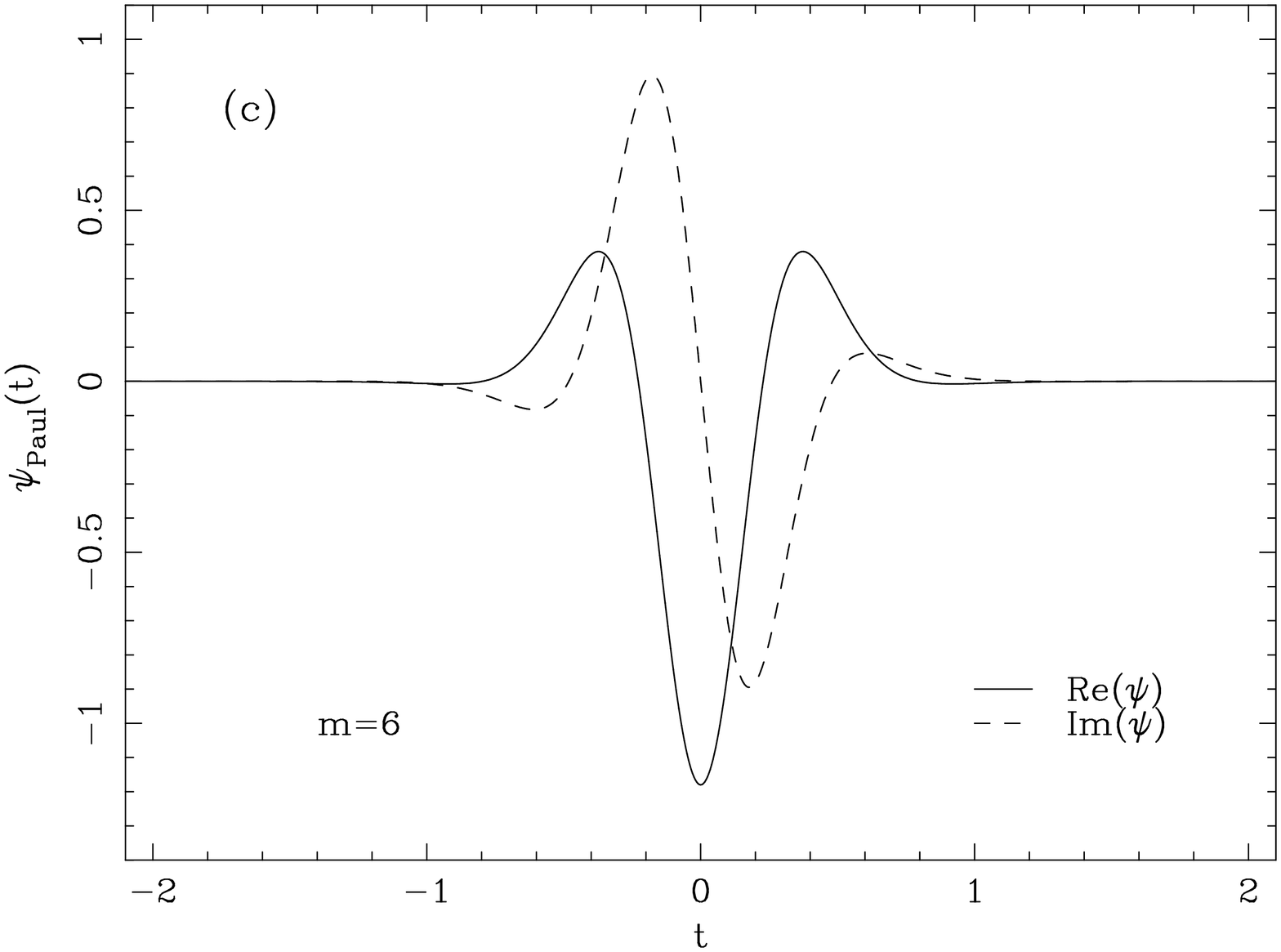}
 \end{center}
 \caption
  {
    Examples of mother wavelets: {\it (a)} Marr wavelet; {\it (b)} Morlet
    wavelet; {\it (c)} Paul wavelet. The real and imaginary parts of complex
    wavelets were plotted by solid and dashed lines, respectively.
  }
 \label{f:falki}
 \end{figure}

There are three most commonly used wavelets in the astronomical time-series analysis: 
{\it (a)} a Marr or Mexican Hat:
\be
   \psi_{\rm Marr}(t) = (1-t^2)e^{-t^2/2},
\label{marr}
\ee
{\it (b)} a Morlet wavelet:
\be
   \psi_{\rm Morlet}(t) = \pi^{-1/4} e^{i2 \pi\nu_0 t} e^{-t^2/2}
\label{morlet}
\ee
which has the complex sinusoidal waveform confined by the Gaussian bell envelope 
(Grossman \& Morlet 1984), and {\it (c)} a Paul wavelet: 
\be
   \psi_{\rm Paul}(t) = \frac{2^mi^m m!}{\sqrt{\pi(2m)!}} (1-it)^{-(m+1)}
\ee
where $m$ denotes an order (Paul 1984; Combes, Grossmann \& Tchamitchian 1989). These 
functions are presented in Fig.~\ref{f:falki}. Application of complex wavelets 
instead of real-valued ones is required to capture information on the 
amplitude and phase simultaneously. On the other hand, a real-valued functions are 
useful for isolation of positive and negative modulations as the separate peaks in the 
time--scale plane, making a clear discrimination of the sharp features and the signal 
discontinuities possible to resolve. 

In our research we are mainly interested in detection and studies of aperiodic flux 
structures, therefore the Morlet wavelet (with $2\pi\nu_0=6$) will be used 
throughout the paper. As it oscillates due to a term of $e^{it}$ and owns a complex form it 
becomes a desired probing tool in tracing and quantifying quasi-periodic modulations at 
different scales.  

\subsubsection{Wavelet power spectrum}
\label{sect:wavelet_power}

For discretely defined time-series $x_k$, $k=0,...,N_{\rm obs}-1$, the wavelet 
transform can be denoted in its discrete form as:
\be
   w_k(a) = \sqrt{\frac{\Delta t}{a}} \sum_{k'=0}^{N_{\rm obs}-1} 
            x_{k'} \psi^\star \left[\frac{(k'-k)\Delta t}{a} \right]
\label{wka}
\ee
where $\Delta t$ is the sampling time of a lightcurve, $k$ represents a localized time 
index and a factor of $\sqrt{\Delta t/a}$ ensures that the wavelet contains the same energy 
everywhere in time--scale space of $w_k(a)$ coefficients. In practice, there is no need to use 
formula (\ref{wka}) for the calculation of wavelet transform. It can be expressed in terms of 
the inverse Fourier transform of the product of the Fourier transforms of the signal and 
wavelet function as given by equation (\ref{a:wkam}) and therefore allow to speed up the 
calculations.

A {\it wavelet power spectrum} can be simply defined as normalized square of the modulus 
of the wavelet transform:
\be
   W = \xi |w_{a,b}|^2 \equiv \xi |w_k(a_m)|^2 .
\ee
A choice of a normalization factor can be done arbitrarily where the most convenient way is 
to set $\xi=\sigma^{-2}$, i.e. equal to the inverse square of the lightcurve variance. Here, 
$\sigma^2$ represents the expectation value of the wavelet transform for a white-noise 
process at each scale $a_m$ and a time location $k$ (TC98). The local wavelet power 
spectrum $W$ with $\xi=\sigma^{-2}$ is distributed as $\chi_2^2$, i.e. as $\chi^2$ 
distribution with two degrees of freedom, therefore it allows to determine the contours at 
given confidence level for every flux oscillation. A reality of any peak in the wavelet map is 
tested against certain background spectrum and thus any peak can be accepted or rejected at 
earlier assumed significance level. 

Due to a finite duration of a lightcurve some artifacts may appear at the edges of the   
wavelet maps. To force them to be negligible one may introduce a {\it cone of influence}
defined as a region of the wavelet map where a wavelet power in the vicinity of signal 
discontinuity decreases by the factor of $e^{-2}$ (TC98).

 \begin{figure}
 \begin{center}
 \includegraphics[width=6cm,angle=-90]{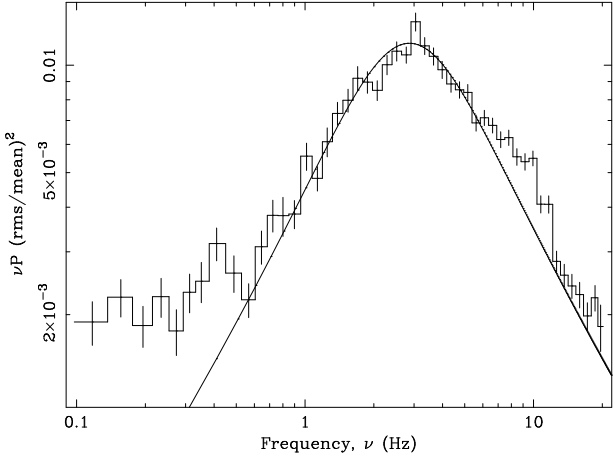}
 \caption
  {
    Fourier power spectrum of Cyg X-1 on Dec 5, 1999. Solid line represents a broad
    Lorentzian peak fitted by Pottschmidt et al. (2003).
  }
 \label{fig:power}
 \end{center}
 \label{f:katja}
 \end{figure}

 \begin{figure}
 \begin{center}
 \includegraphics[width=6cm,angle=-90]{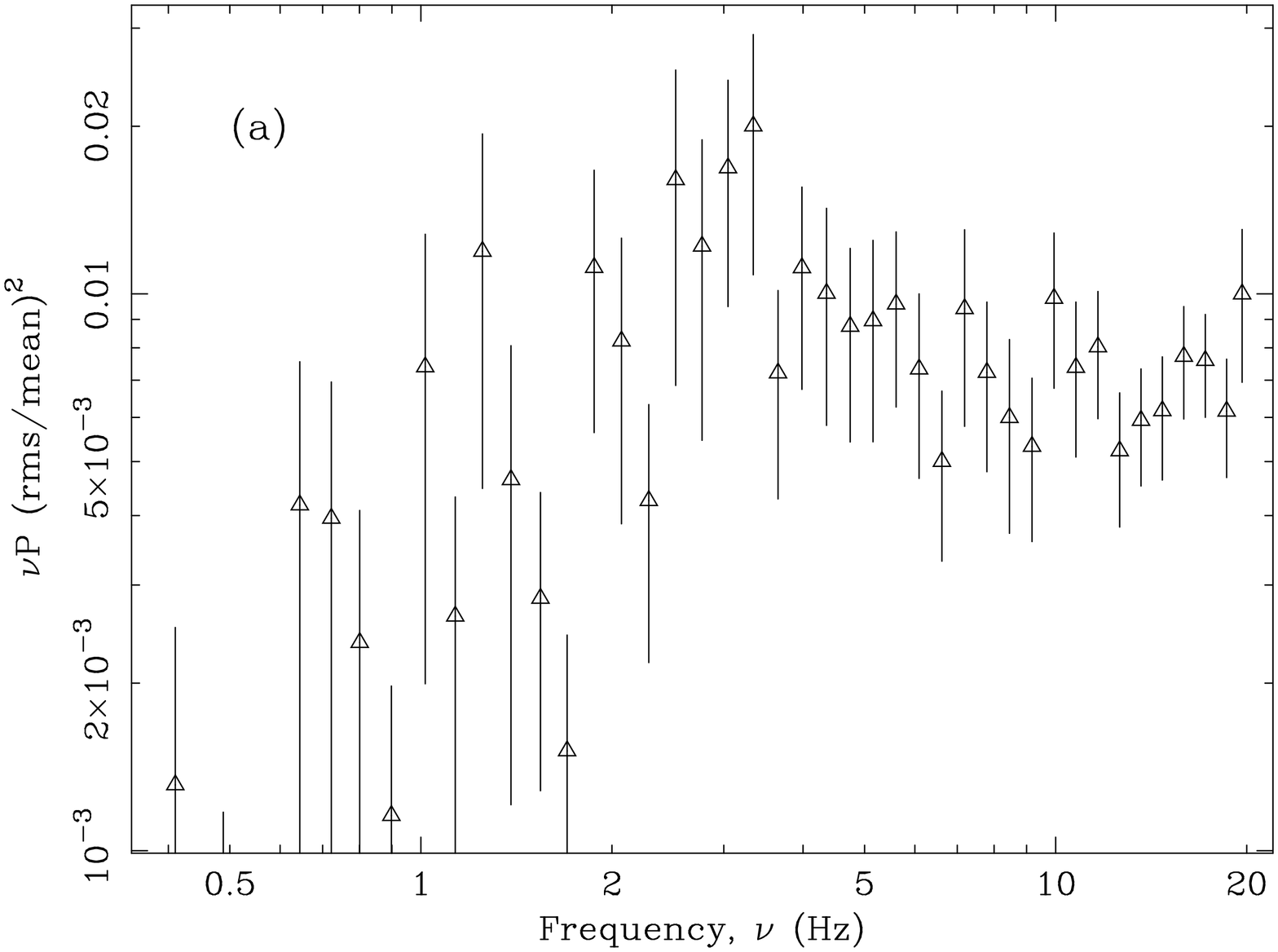}
 \includegraphics[width=6cm,angle=-90]{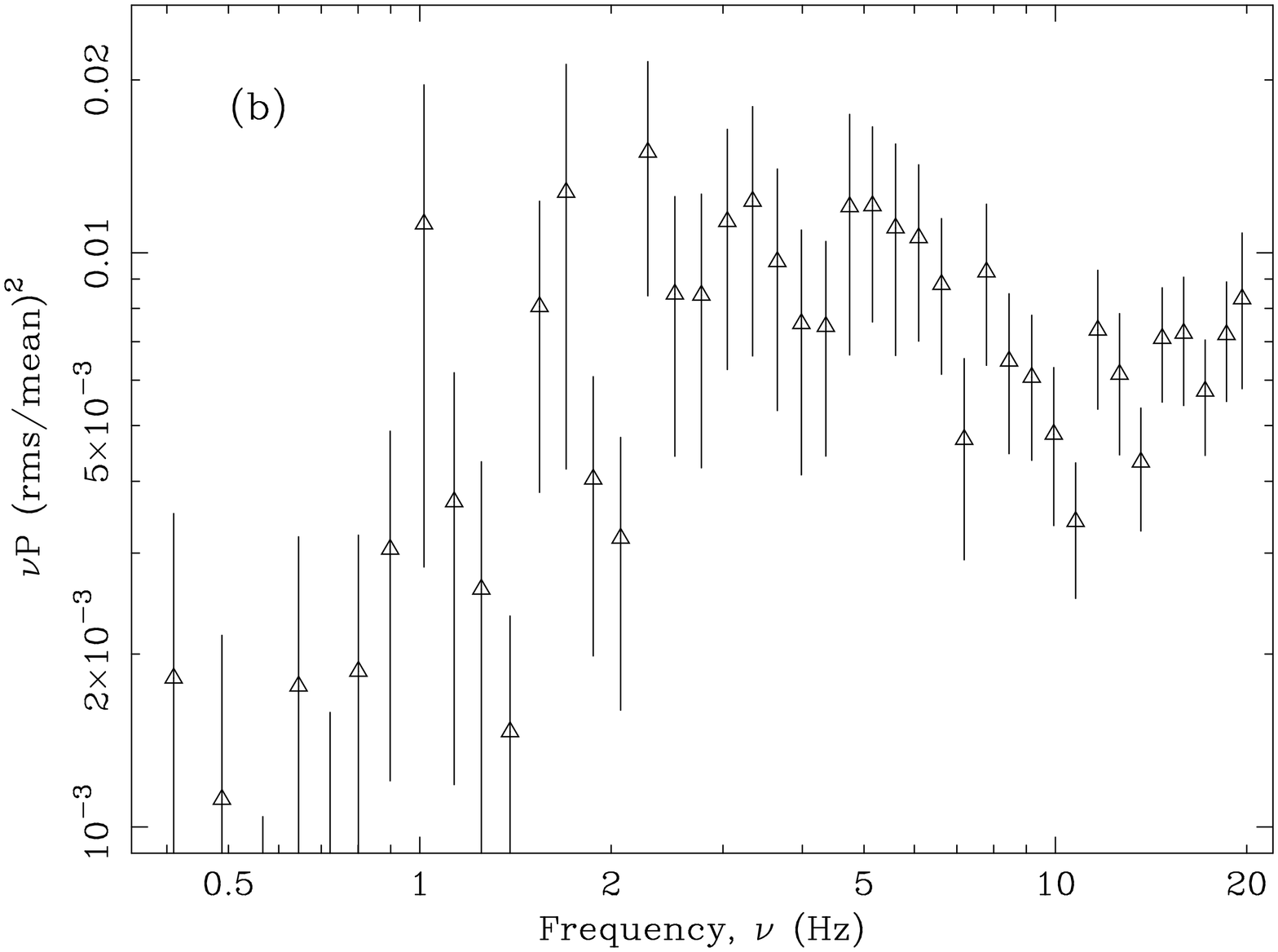}
 \includegraphics[width=6cm,angle=-90]{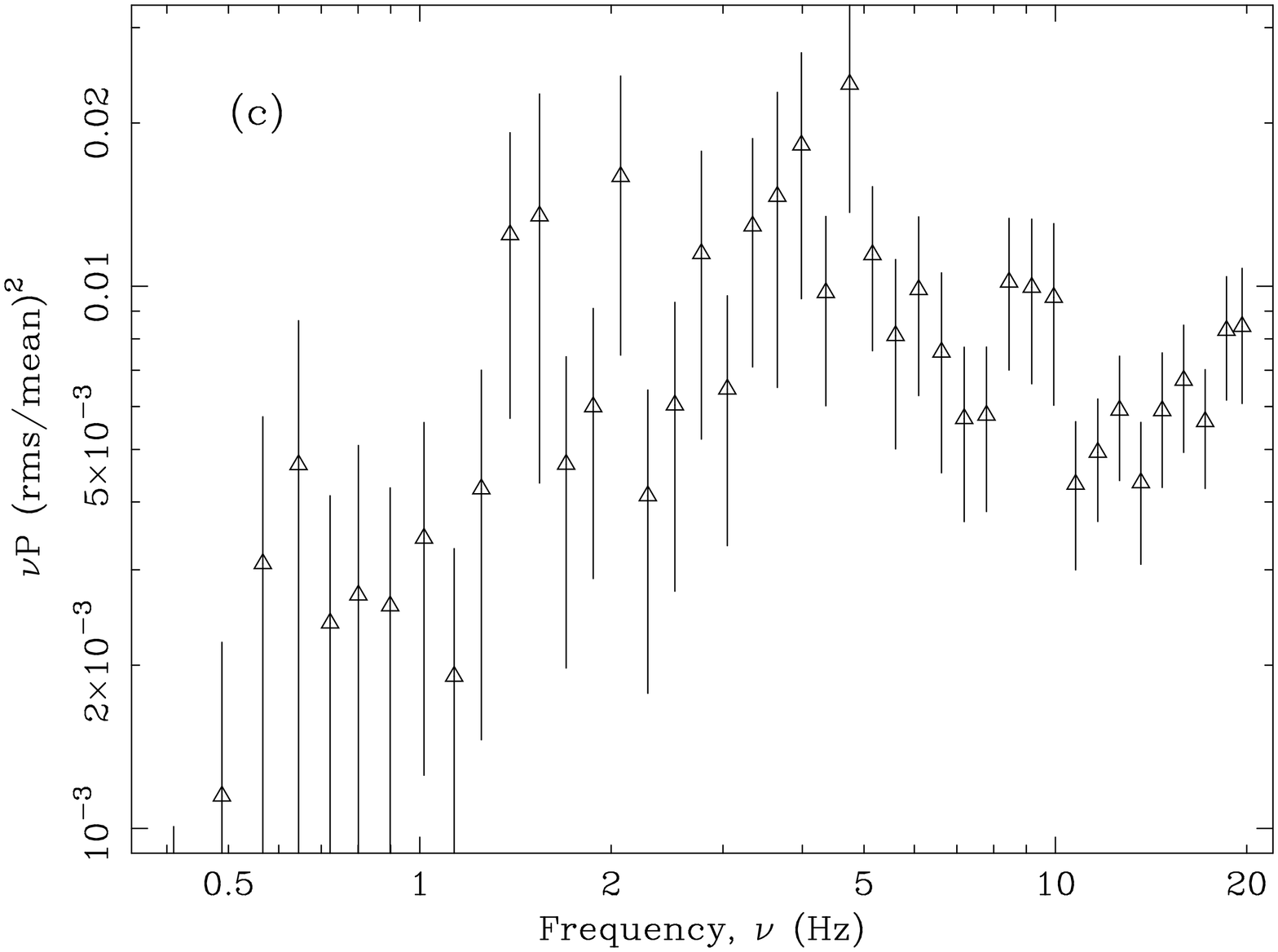}
 \caption
  {
    Exemplary power spectra calculated for the single consecutive data streams of duration
    20 s corresponding to the PCA lightcurve intervals: {\it(a)} 1090--1110 s, {\it(b)}
    1110--1130 s, {\it(c)} 1130--1150 s. For purposes of a comparison with the global wavelet
    power spectra in Fig.~\ref{fig:cygx1_v6}, the subtraction of a noise level was not
    applied.
  }
 \end{center}
 \end{figure}

2-D information about the signal in the wavelet map also allows us to compare it to the 
results of Fourier power spectrum. It can be achieved by performing an integration of 
the wavelet spectrum over time
which is known under the name of a {\it global wavelet (power) spectrum} or a {\it scalegram} 
technique (Scargle et al. 1993). Its discrete form can be denoted as:
\be
    G(a_m) = \frac{\xi}{N_{\rm obs}} \sum_{k=k_1}^{k_2} |w_k(a_m)|^2
\label{gam}
\ee
where $0\leq k_1<k_2 \leq N_{\rm obs}-1$ and $\xi$ denotes the normalization factor. Here, 
$k_1$ and $k_2$ ought to be chosen to lie outside the cone of influence for particular 
scale $a_m$. Since the normalization can be arbitrary, one can choose it the same as for the 
wavelet power spectrum (e.g. $\xi=\sigma^{-2}$) and therefore determine the proper 
significance levels (see TC98 for details). If so, when a comparison of (\ref{gam}) with the
calculated Fourier power spectrum is required, it is possible to adopt a
normalization factor of the wavelet spectrum $\xi=2\Delta t/\bar{x}^2$. Thus, the global 
wavelet power spectrum will have the units of (rms/mean)$^2$/Hz.

A relation between the wavelet scale $a$ and the Fourier frequency $\nu$ depends on our 
choice of the wavelet function. For the Morlet function ($2\pi\nu_0=6$) it holds $\nu_{\rm 
Morlet}\simeq (1.03a)^{-1}$, for the Marr wavelet $\nu_{\rm Marr}\simeq (3.97a)^{-1}$ whereas 
for the Paul wavelet ($m=6$) $\nu_{\rm Paul}\simeq (0.97a)^{-1}$. Hereafter, y-axes for all of our 
wavelet maps will be expressed as frequency, given by the above transformations.


\section{Results}
\label{s:s4}

 \begin{figure*}
 \begin{center}
 \includegraphics[width=13cm,angle=0]{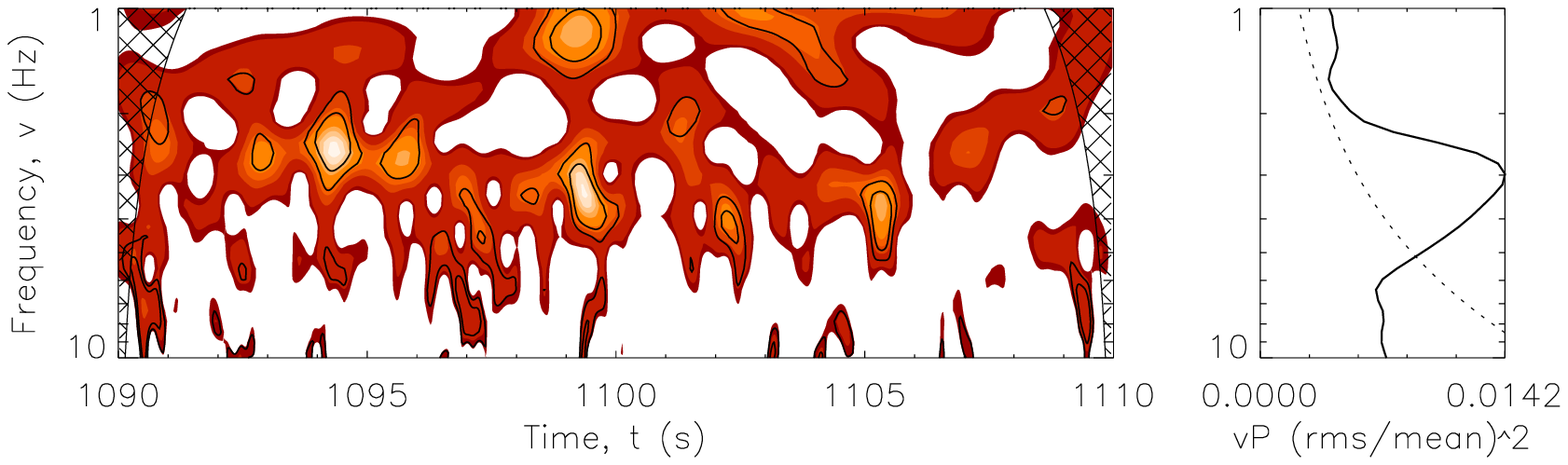}
 \includegraphics[width=13cm,angle=0]{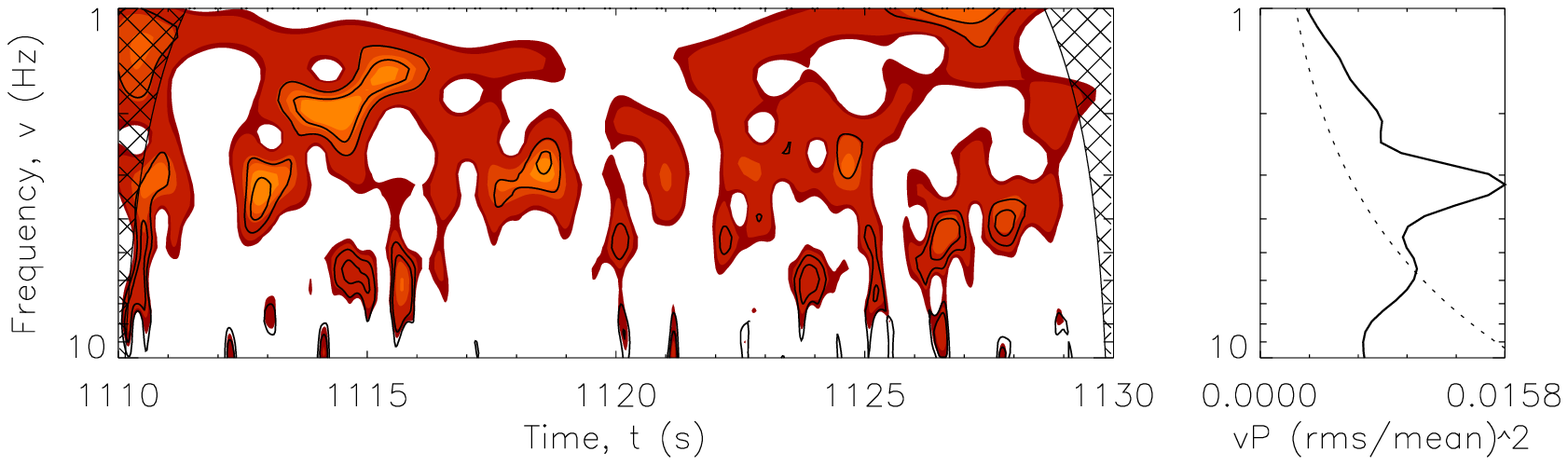}
 \includegraphics[width=13cm,angle=0]{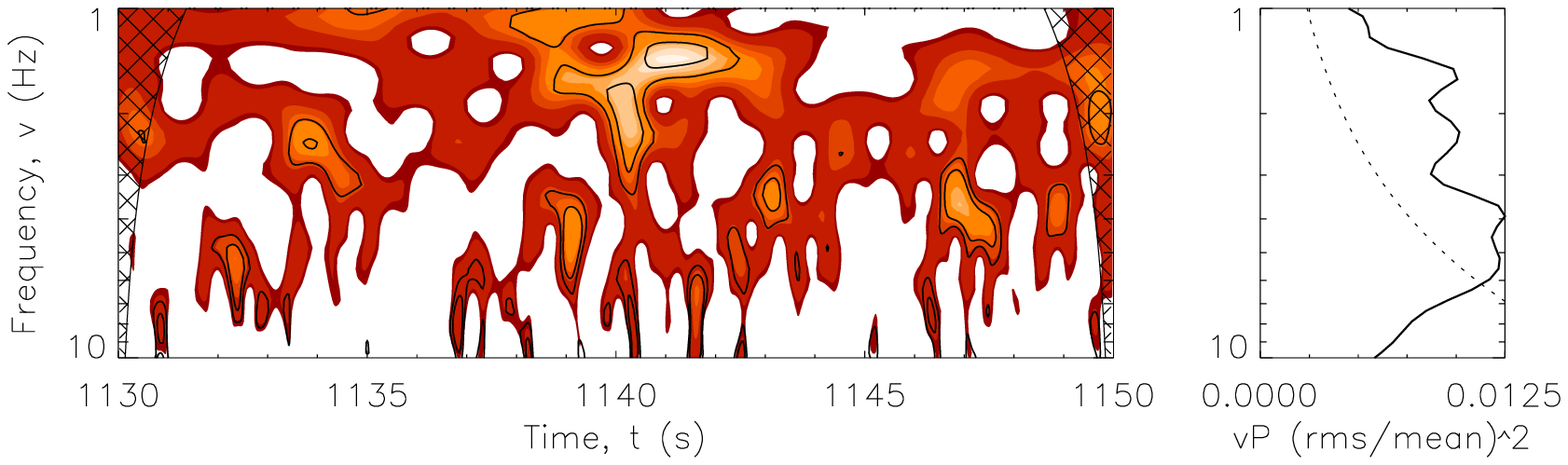}
 \caption
  {
   {\it (Left panels)}
   Wavelet power spectra of Cyg X-1 during its failed state transition on 1999.12.05
   mapping a time--frequency evolution of $\sim3$ Hz quasi-periodic feature (calculated for
   randomly selected 60 s time interval from the PCA time-series). Top values of wavelet
   power are denoted by gradual brightening of the color. Black solid contours denote
   significance level of 90\% (outer) and 99\% (inner) for detected peaks. 3-D visualization of 
   the 1090--1110~s
   wavelet map is shown in Fig.~\ref{fig:3D}. A cone of influence is marked as dashed region.
   {\it (Right panels)}
   Corresponding global wavelet power spectra, drawn as integrated over time wavelet power
   spectrum multiplied by the frequency (see Fig.~4 for a comparison with the
   Fourier PSD of the same time sequences). Dotted lines represent the 95\% significance
   level calculated according to the procedure given in TC98.
  }
 \label{fig:cygx1_v6}
 \end{center}
 \end{figure*}

The selected data set covers 1364 s. In Fig.~3 we show the corresponding PSD, together 
with the dominant Lorentzian component fitted by P03. The shape of 
the Lorentzian component was described as:
\be 
    L_2(\nu )=\pi^{-1} \frac{2R_2^2Q_2\nu_2}{\nu_2^2+4Q_2^2(\nu-\nu_2)^2} ,
\ee
and the best fit values of the parameters were: amplitude $R_2=0.188\pm 0.004$, center 
frequency $\nu_2=1.769^{+0.066}_{-0.070}$~Hz, and the quality factor $Q_2= 
0.395^{+0.032}_{-0.035}$. This power spectrum component has a maximum at the frequency 
2.856~Hz, and a characteristic decay time-scale, $ \tau_2 = Q_2/(\pi \nu_2)$, is 
given by $\tau_2=0.07$~s.

The presented PSD shows relatively small errors since the whole time sequence
was used. The whole lightcurve was divided into intervals of 25.6 s, the PSD
was calculated for each part separately, and finally
averaged PSD was plotted, as customary done for such data.

When single PSD spectra are not averaged, they show a considerable scatter. This 
scatter is expected from the $\chi^2$ distribution of the individual PSD values.
For the purpose of further discussion of the wavelet maps, we show (see Fig.~4)
the examples of the PSD obtained for the single fragments of the lightcurve of duration
20~s. Errors are large but the Lorentzian shape is still visible. The selection of
60 s long sequence from whole duration of our PCA time-series was done by us randomly,
here corresponding to the interval between 1090--1150 s (Fig.~\ref{fig:krzywa}).

\subsection{Wavelet maps}
\label{ss:wamaps}

We construct the wavelet maps for the same parts of the lightcurve as used in the PSD analysis. 
The selected band of scales corresponds to the Fourier frequency range between 1 and 10 Hz, 
where the Lorentzian has the maximum.

We use the Morlet wavelet, with the standard assumption $2 \pi \nu_0 = 6$ 
(see Section~\ref{sect:wavelets}). This means that we probe the signal with
a damped wave performing $\sim 3$ oscillations. 

The result is shown in Fig.~\ref{fig:cygx1_v6}, for three consecutive parts of
the lightcurve. Integrated wavelet spectra are shown to the right of each 
of the sequences. 

 \begin{figure*}
 \begin{center}
 \includegraphics[width=9.6cm,angle=0.0]{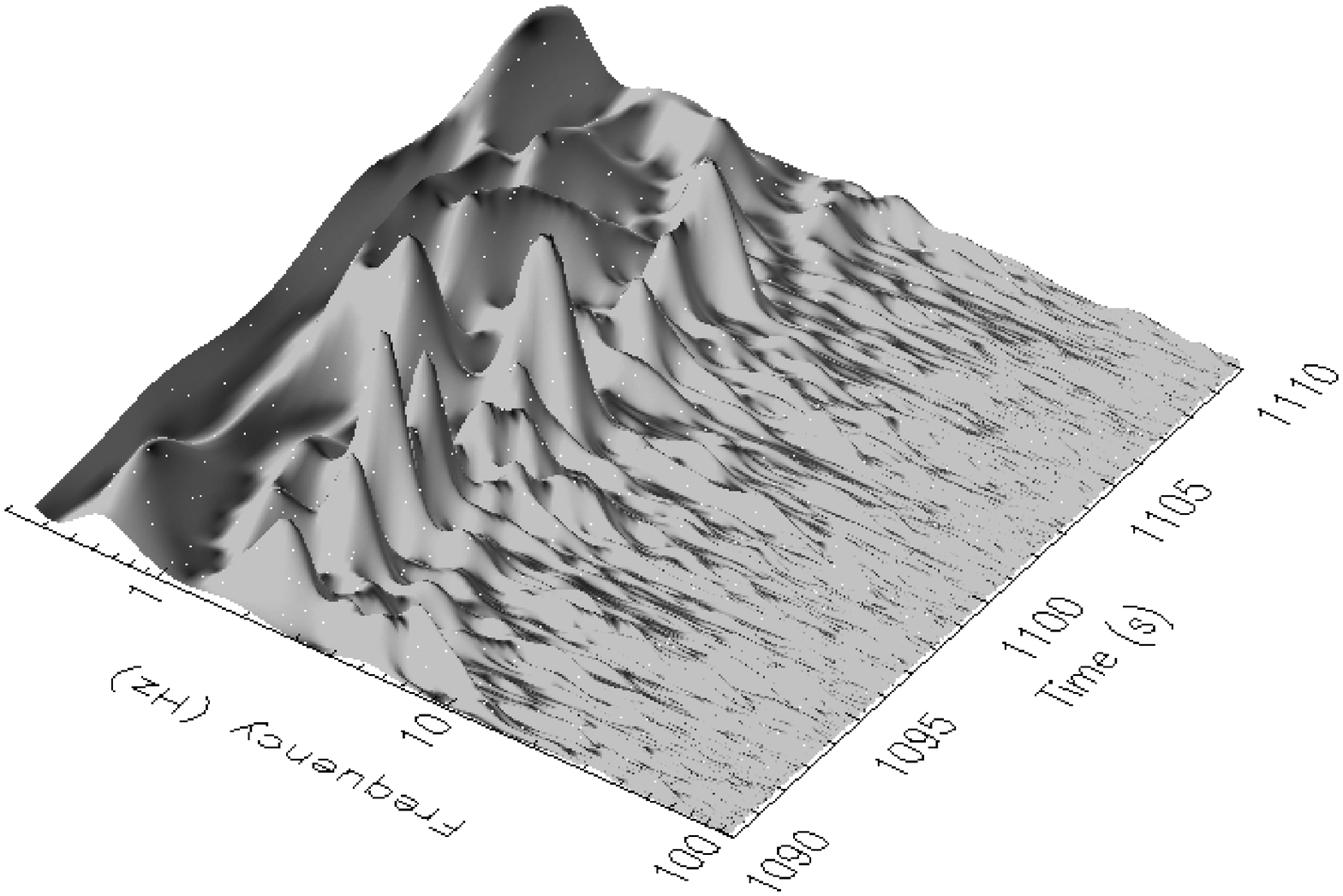}
 \includegraphics[width=9.6cm,angle=0.0]{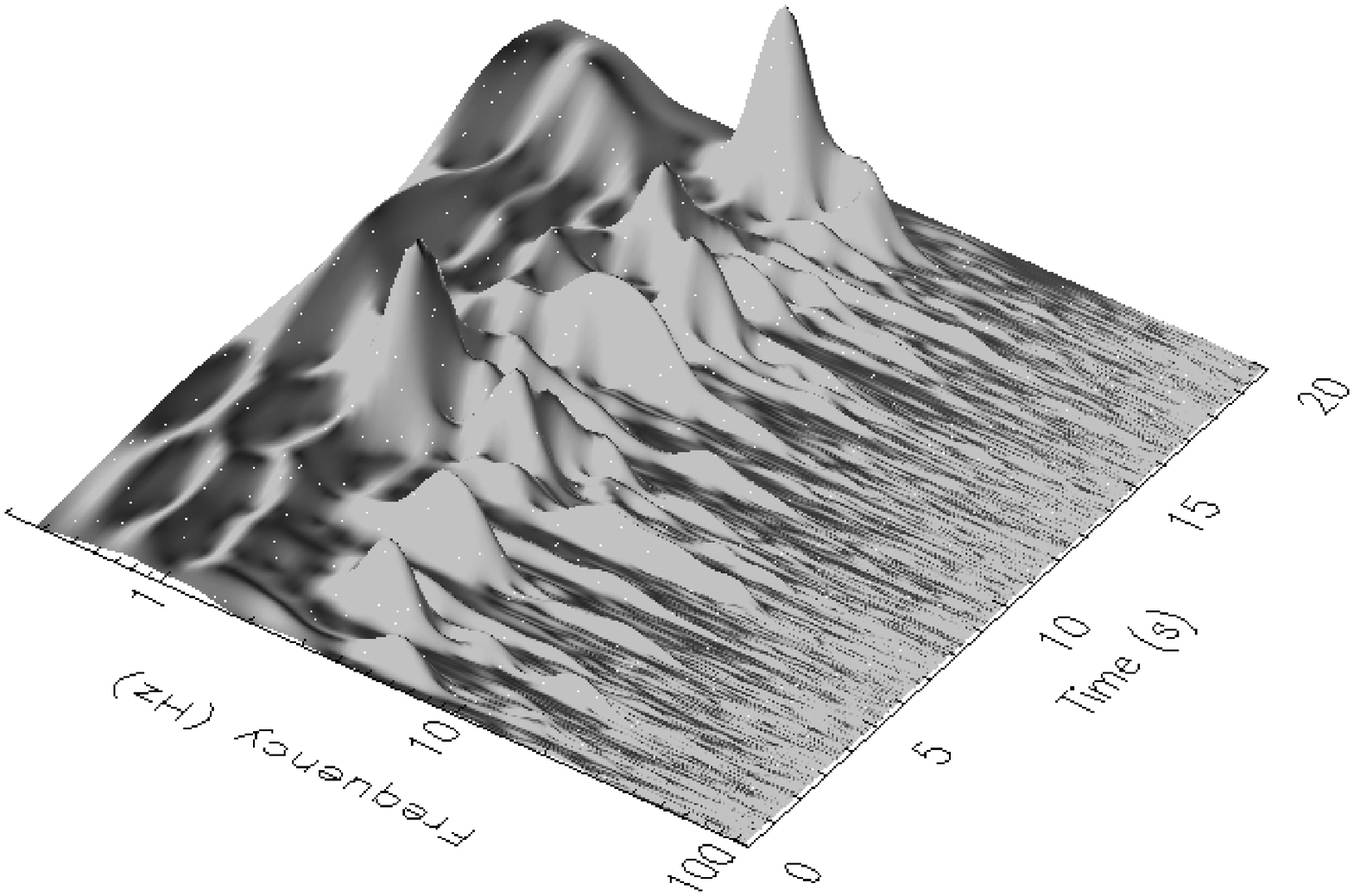}
 \caption{
          3-D presentation of the wavelet power spectrum: {\it (upper)} for 20 s long PCA
          lightcurve; {\it (bottom)} for 20 s segment from a simulated
          lightcurve containing randomly occurring flares characterized by the Lorentz profile
          (see Section \ref{ss:simlc}). In order of visualization of wavelet power
          distribution up to 100 Hz, maps were calculated for the lightcurves with the bin time
          of $\Delta t = 2^{-9}$ s.
         }
 \label{fig:3D}
 \end{center}
 \end{figure*}

 \begin{figure*}
 \begin{center}
 \includegraphics[width=13cm,angle=0]{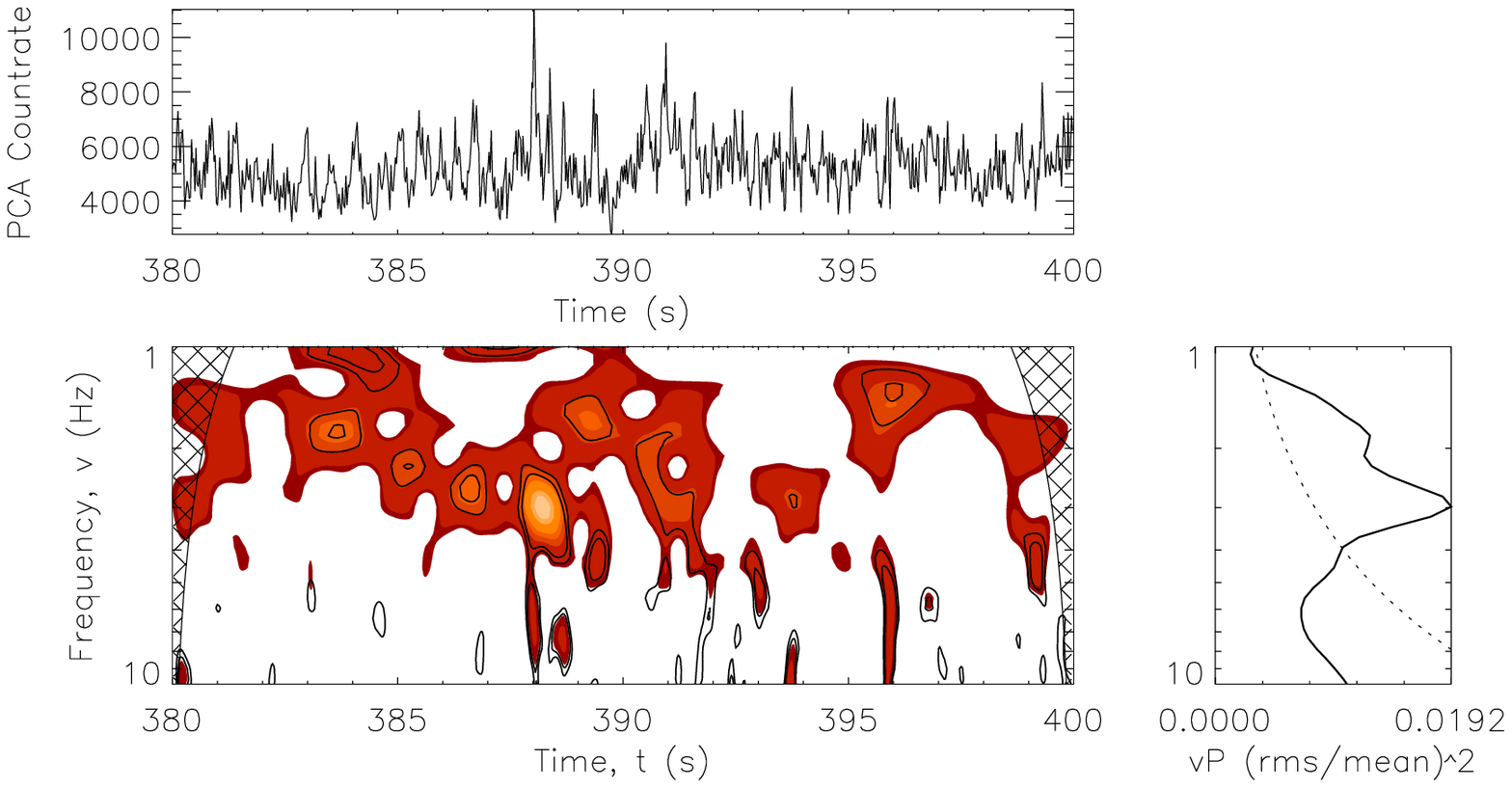}
 \caption
  {
   {\it (Central panel)} Wavelet power spectrum of Cyg X-1 revealing a puzzling chain of peaks
   ranging between 383--391 s of the analyzed PCA lightcurve {\it (top panel)}. In the right,
   corresponding global wavelet power spectrum. See Section \ref{ss:trace} for discussion.
  }
 \label{fig:5islands}
 \end{center}
 \end{figure*}

In all maps many localized strong peaks are seen, standing clearly out of the background. 
This is even better seen in a 3-D plot (see Fig.~\ref{fig:3D}, upper panel). Those peaks 
dominate the integrated spectrum although they are present only occasionally, mostly 
between 2--5 Hz. Many lower but still significant peaks are also present, so active phase 
with significant peaks covers more than half of the observed time. No single frequency 
seems to be favored and we do not see any particular evolutionary trend. The largest 
peaks extend in time typically for $\sim 1$~s which means that they are practically 
unresolved.  Those which extend for a few seconds can be elongated either in the 
direction of lower or higher frequencies, and the shapes are rather irregular.

\begin{table}
\caption{Properties of the localized peaks in the wavelet maps for the assumed significance
level of 90\%. 
}
\label{tab1}
\begin{tabular}{ccc}
\hline
Sequence &  Cyg X-1  & Simulated lightcurve (equation \ref{eq:lorentz_przepis}) \\
\hline
\multicolumn{3}{c}{\it Number of localized peaks}\\
Seq. 1  & 18   & 11  \\
Seq. 2  & 16   & 16  \\
Seq. 3  & 17   & 12  \\
\multicolumn{3}{c}{\it Activity level, $\kappa$, in \%}\\
Seq. 1  & 63.6 & 51.2  \\
Seq. 2  & 59.9 & 59.5  \\
Seq. 3  & 62.3 & 67.9  \\
\multicolumn{3}{c}{\it Median of peak duration, $\Delta T_p$, in {\rm [s]}}\\
Seq. 1  & 0.54 & 0.54  \\
Seq. 2  & 0.54 & 0.45  \\
Seq. 3  & 0.27 & 0.54  \\
\multicolumn{3}{c}{\it Median of peak width, $\Delta \log \nu_p $}\\
Seq. 1  & 0.12 & 0.13  \\
Seq. 2  & 0.14 & 0.11  \\
Seq. 3  & 0.14 & 0.15  \\
\multicolumn{3}{c}{\it Median of peak frequency, $\nu_p$, in {\rm [Hz]}}\\
Seq. 1  & 3.7  & 3.9 \\
Seq. 1  & 4.3  & 3.8  \\
Seq. 3  & 5.2  & 2.8 \\
\multicolumn{3}{c}{\it Peak width, $\Delta \log \nu_h $, of the highest peak}\\
Seq. 1  & 0.25 & 0.23  \\
Seq. 2  & 0.21 & 0.35  \\
Seq. 3  & 0.39 & 0.47  \\
\multicolumn{3}{c}{\it Peak duration, $\Delta T_h$, of the highest peak in {\rm [s]}}\\
Seq. 1  & 1.1  & 1.2  \\
Seq. 2  & 2.8  & 1.6  \\
Seq. 3  & 2.7  & 2.7  \\
\multicolumn{3}{c}{\it Peak frequency, $\nu_h$, of the highest peak in {\rm [Hz]}}\\
Seq. 1  & 3.4  & 2.2 \\
Seq. 2  & 2.1  & 1.7  \\
Seq. 3  & 1.4  & 1.5  \\
\hline
\end{tabular}
\end{table}

We analyzed the distribution and the properties of the significant peaks
present in the observational data. The results are given in Table~\ref{tab1}.
For the analysis, we took only those peaks which are well determined, i.e.
detected with 90\% confidence level and not extending beyond the studied frequency
range. We give there the number of such localized peaks in each of the three
sequences. We define the activity level, $\kappa$, in each sequence as the fraction of
time when at least one peak at any frequency is present. We determine the median value of
the logarithm of the frequency, $\Delta \log \nu_p$, the median of the peak
duration, $\Delta T_p$, and the median of the peak frequency, $\nu_p$.
We also choose the highest of the peaks in each sequence and give for such
peak its duration, $\Delta T_h$, frequency $\nu_h$, and uncertainty of its
localization, $\Delta \log \nu_h$.

We see that most of the peaks are practically unresolved since their duration
and the uncertainty of the frequency is comparable to the Heisenberg limits
of the wavelet analysis. For the Morlet wavelet
with $2 \pi \nu_0 = 6$ the minimum value of $\Delta \log \nu = 0.11$, and
the minimum value of the time resolution is 0.24 s at 3 Hz (0.71 s at 1 Hz and
0.07 s at 10 Hz). The count rate of the source is high ($\sim 6000$ cts~s$^{-1}$)
which allows to reach the formal limit of the adopted approach.

Searching the whole lightcurve we have found several interesting sequences. 
One of them is shown in Fig.~\ref{fig:5islands}. There seems to be a chain of peaks,
systematically moving towards higher frequencies. Such a development
could be consistent with propagation of the perturbations towards the gravity
center, with the expected time-scales decreasing with the radius. We discuss it
in Section \ref{ss:trace}.

For the purpose of a comparison, the contours corresponding to 99\%  
significance level were drawn additionally on the wavelet maps. We found that 
at this level of significance only the most prominent features of the local 
time--frequency variability are still detected. They cover a relatively small
fraction of time in comparison to 90 \% contours, with larger differences between
the consecutive time sequences. Therefore, it seems that 90\% contours are more 
convenient for characterizing the activity level, $\kappa$, of the lightcurve, 
whereas 99\% contours should be used in verification of any trends in the frequency 
domain.


\section{Discussion}

The nature of the short time-scale variability in X-ray emission of accreting 
binaries is still under discussion. The PSD is generally broad-band so
the variability is successfully modeled as a shot noise (Terrell 1972;
Lehto 1989). Later, more specific models with some physical background were
developed (disk turbulence, Nowak \& Wagoner 1995; magnetic flares in the disk
corona, Poutanen \& Fabian 1999, \. Zycki 2002; propagation of perturbations 
in the inner hot flow, B\" ottcher \& Liang 1999, \. Zycki 2003; 
perturbations in the accretion rate in the cold disk, Mineshige, Ouchi \&
Nishimori 1994, Lyubarskii 1997). Differentiation between those models
is very difficult.

Occasionally, the power spectrum shows a hint of a single oscillation which dominates
the variability. Such a situation happens during a transition state. Careful study of 
this apparently simple state should reveal the nature of variations more easily. For
example, during a failed transition state, the PSD of Cyg X-1 was quite well represented 
by a single Lorentzian peak (P03).

Such a single Lorentzian peak represents a damped coherent oscillation with a  
time-independent frequency. This damping is responsible for the width of the peak. 
However, similarly broad power spectrum can in principle result from oscillations which 
are not so strongly damped but which change the frequency with time, like in case of a 
chirp signal (see e.g. Cohen 1995). Such a change in oscillation frequency may be, for 
example, connected with the accretion process: if the wave or hot plasma moves inward, 
the local characteristic scales may decrease with radius.

The Fourier analysis cannot differentiate between the damped oscillation and the modulated 
frequency. Therefore, in order to search for the underlying physical process we need more 
advanced method of the analysis.

In this paper we applied wavelet analysis to the Cyg~X-1 lightcurve. The wavelet
maps show that strong, well localized oscillations are present rather occasionally, 
at frequencies $\sim$ 3--5 Hz and they typically last for $\sim 0.3 - 0.5$ s. A few such
strong peaks are always present in each of the 20 s sequences and they 
effectively determine the character of variability, although some peaks above the $90\%$ 
confidence threshold are typically seen for more than half of the observation time. 
We did not find any systematic frequency change in oscillations. The analysis gave an 
upper limit on the rate of the frequency change during a single oscillation $\delta\ln 
\nu_0/\delta\ln t \le 0.2$.

This result supports the option of a coherent oscillation with strong damping present in 
the system. However, in order to analyze more qualitatively the description of the system 
as a damped oscillator we performed numerical simulations.

\subsection{Lightcurves simulated in time domain}
\label{ss:simlc}

In order to model the observed lightcurve dominated by the Lorentzian peak we basically 
follow the approach of \. Zycki (2003). We create a simulated lightcurve as a 
superposition of the random shots with the shape:
\be
F(t) = A_i e^{-(t-t_i)/\tau} \cos[2\pi \nu_2 (t-t_i) + \phi_i] ~~{\rm for}~ t>t_i,
\label{eq:lorentz_przepis} 
\ee
and $F(t) = 0$ for $t<t_i$. The values of the frequency $\nu_2$ and of the characteristic 
time-scale $\tau$ were fixed, i.e. the same for all shots. The amplitude of a shot, $A_i$,
was assumed to cover uniformly the range between 0 and $A_0$, the shot phase, $\phi_i$, 
was assumed to be distributed uniformly between 0 and $2 \pi$. The moments of shot 
appearance, $t_i$, were chosen from the Poisson distribution, for an assumed mean 
frequency of shot generation, $\lambda$. It means that the time separation between the 
consecutive flares, $\delta t =t_{i+1}-t_i$, was given by $\delta t = -\log(rand)/\lambda$ 
where $rand$ is a random number between 0 and 1.

 \begin{figure}
 \begin{center}
 \vspace*{5pt}
 \includegraphics[width=6cm,angle=-90]{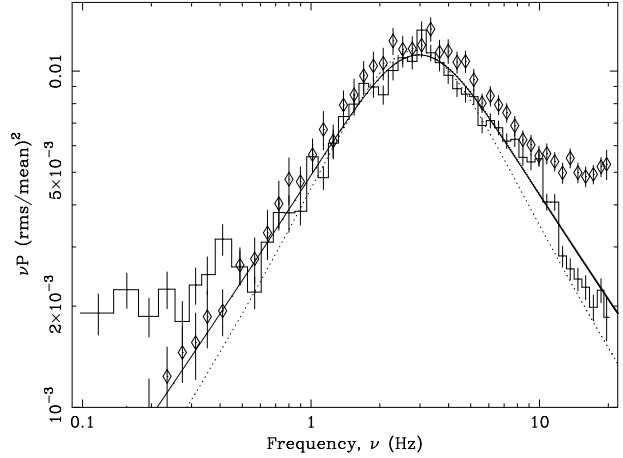}
 \caption
  {
   Comparison of the fits to the PCA power spectrum performed by Pottschmidt et al. (2003)
   ({\it dotted line}) and provided by the application of equation (\ref{eq:two_lorentz})
   ({\it solid line}). {\it Diamond} markers denote power spectrum calculated from 1000 s long
   simulated lightcurve (see Section \ref{ss:simlc}).
  }
\label{fig:lorentz}
 \end{center}
 \end{figure}

It is easy to show that such a lightcurve is characterized by a power density spectrum:  
\be
  {\cal P}(\nu) = \frac{C}{\tau^{-2}+4\pi^2(\nu-\nu_2)^2}+\frac{C}{\tau^{-2}+4\pi^2(\nu+\nu_2)^2}
\label{eq:two_lorentz}
\ee
The first term has the Lorentzian shape, as requested. The second term is a 
smooth and slowly varying function so the sum of these two terms roughly preserves
the Lorentzian shape. We cannot have a better representation of
a Lorentzian power spectrum since the power spectrum of a real function in the time 
domain must have a power spectrum symmetric in the frequency. The Lorentzian shape
itself is not symmetric so it corresponds to unphysical complex function in the
time domain. 

This complication means that formally we cannot use the published values of
the Lorentzian peak's parameters in order to generate the simulated lightcurves
with the same power spectrum. Instead, we should refit the original power 
spectrum with the new function given by equation~(\ref{eq:two_lorentz}). Since this 
shape is not strongly different from a Lorentzian (see Fig.~\ref{fig:lorentz}), 
it represents the data equally well. The new values of the constants involved
are: $C=0.945 \pm 0.065$, $\nu_2=1.817 \pm 0.184$~Hz, $\tau=0.067 \pm 0.003$~s.  
New value of the frequency and the damping time-scale are only slightly different from 
the values obtained by P03 from fitting a single Lorentzian peak 
(1.769 Hz and 0.07 s, correspondingly).

 \begin{figure*}
 \begin{center}
 \includegraphics[width=13cm,angle=0]{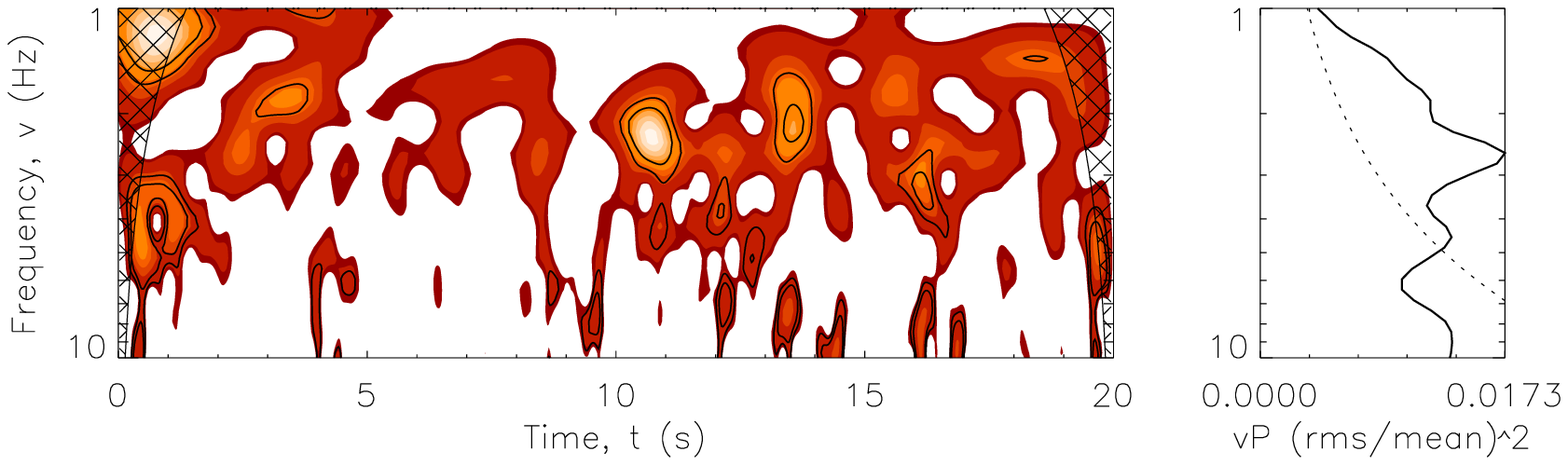}
 \includegraphics[width=13cm,angle=0]{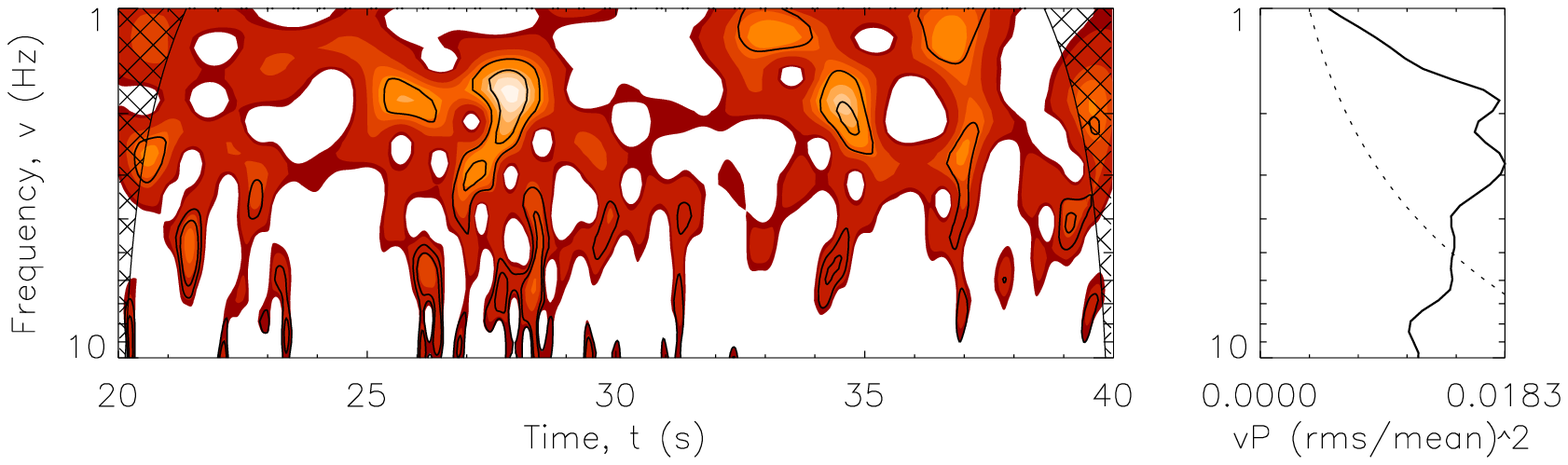}
 \includegraphics[width=13cm,angle=0]{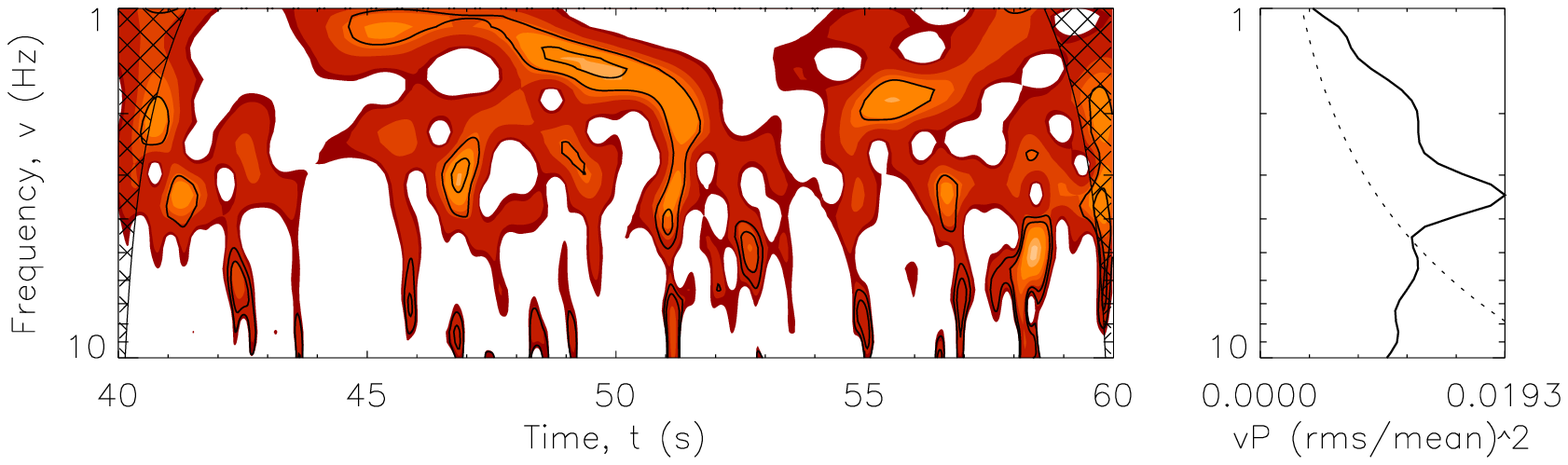}
 \caption
  {
   {\it (Left panels)}
   Wavelet power spectra of the simulated lightcurve ($\lambda=20$, $A_0=2500$ cts~s$^{-1}$)
   calculated for randomly
   selected 60-second time interval. Top values of the wavelet power are denoted by
   gradual brightening of the color. Black solid contours denote significance level of 90\%
   (outer) and 99\% (inner)
   for detected peaks. A cone of influence is marked as dashed region.
   {\it (Right panels)}
   Corresponding global wavelet power spectra, drawn as integrated over time wavelet power
   spectrum multiplied by the frequency. Dotted lines represent the 95\% significance level.
  }
\label{fig:sim_v6}
 \end{center}
 \end{figure*}

The lightcurve given by equation~(\ref{eq:lorentz_przepis}) has statistically zero mean 
so we must add the value representing the mean count rate, $\bar{x}$, in the PCA observation.
The coefficients $A_0$ and $\lambda$ must be chosen in such way that the resulting 
lightcurve has the required rms, the same as PCA data. Interestingly, we found that there 
exists a whole family of parameters ($A_0$,$\lambda$) which reproduce the required 
normalization of the power spectrum since the constant $C$ in equation 
(\ref{eq:two_lorentz}) is given by:
\be
   C \propto \frac{A_0^2 \lambda}{\bar{x}^2}
\ee
what implies that the following relation:
\be
   A_0^2 \lambda = \text{const}
\label{eq:Alambda}
\ee
holds for every couple of ($A_0$,$\lambda$) at assumed mean count rate, $\bar{x}$, and 
fractional rms variability.

 \begin{figure*}
 \begin{center}
 \includegraphics[width=12.9cm,angle=0]{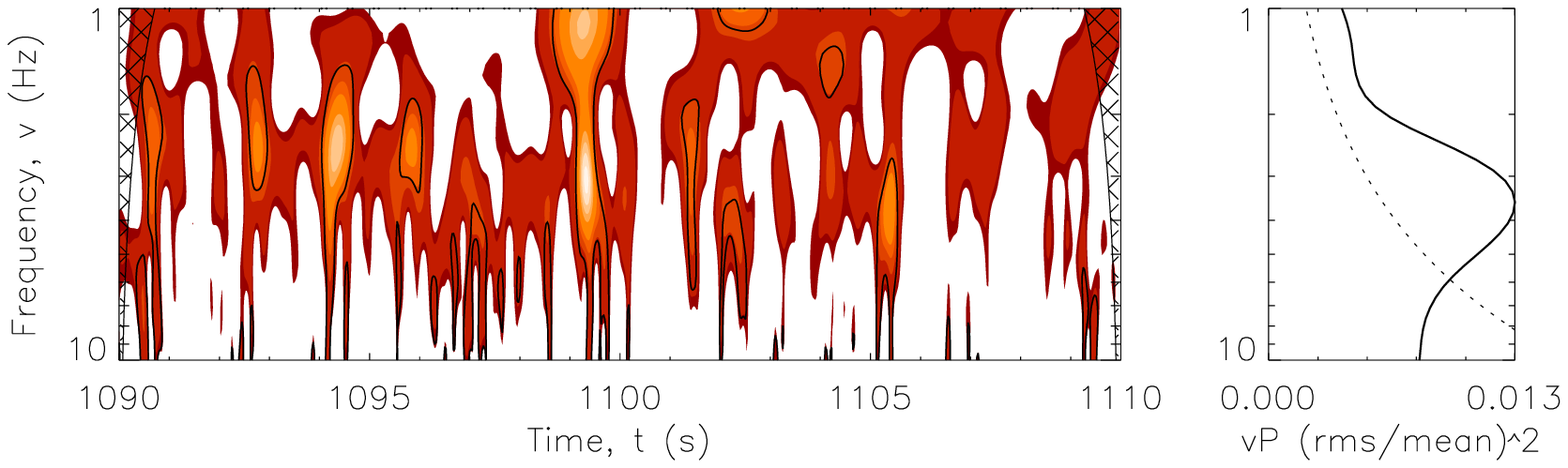}
 \includegraphics[width=12.9cm,angle=0]{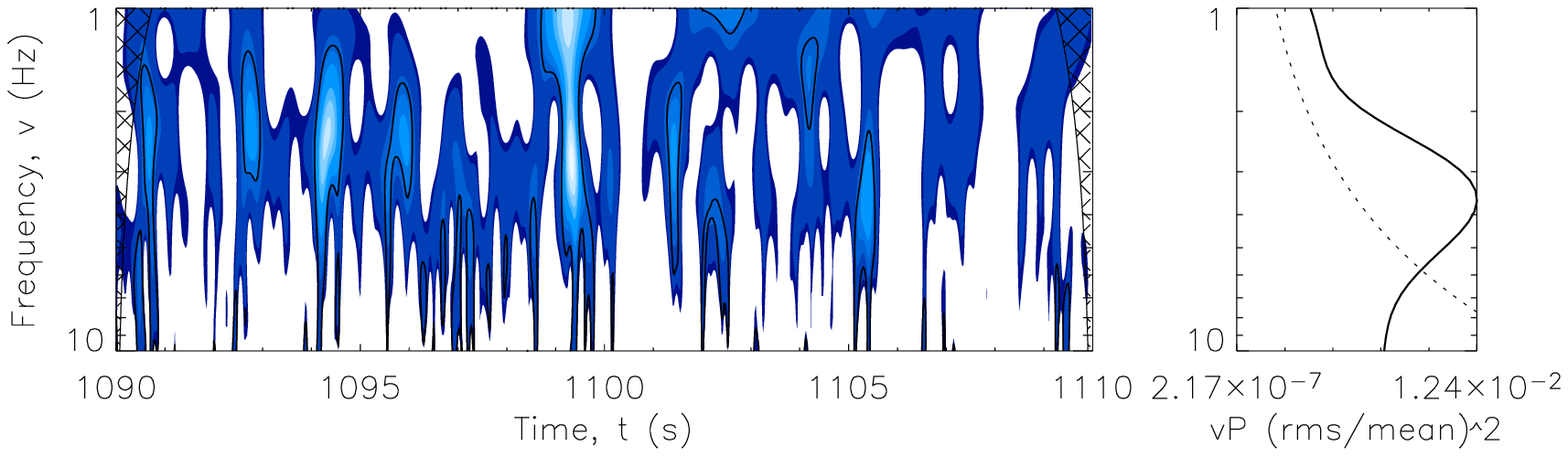}
 \includegraphics[width=12.9cm,angle=0]{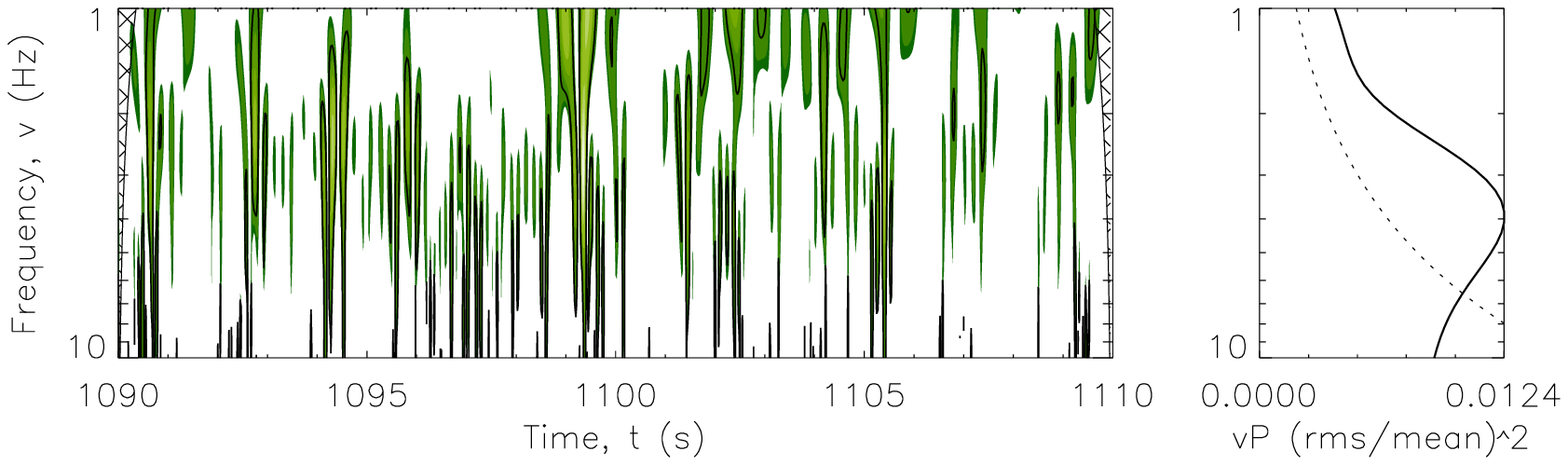}
 \caption
  {
   Wavelet power spectra of Cyg X-1 corresponding to the PCA data sequence between
   1090--1110 s, calculated applying {\it(top panel)} Morlet wavelet with $2\pi\nu_0=3$,
   {\it(middle panel)} Paul wavelet with $m=6$ and {\it(bottom panel)} Marr wavelet.
   In the right, corresponding global wavelet power spectra.
  }
\label{fig:inne}
 \end{center}
 \end{figure*}

We also included the white noise in our simulated lightcurve. For this purpose, in each 
of the time bins of the simulated time-series we calculate the number of photons, we treat 
this value as 'average', draw an actual number of counts from the Poisson distribution 
around this value and finally we divide this value by the bin size in order to return to 
count per second units. It can be shortly written as:
\be
   F_N(t) = \mbox{\sc poisdev}[F(t)\Delta t]/\Delta t .
\ee
Above, $F(t)$ is given by equation (\ref{eq:lorentz_przepis}), {\sc poisdev} represents
a subroutine which generates a random number with the Poisson distribution (Press et al. 1992)
and thus $F_N(t)$ denotes simulated lightcurve with the Poisson noise level included.

The level of noise does not depend on a specific choice of a pair ($A_0$, $\lambda$) if it 
satisfies the relation (\ref{eq:Alambda}). However, created lightcurves differ visually: low 
$\lambda$s naturally result in rare steep peaks while high $\lambda$s give 
apparently very noisy curve without clear pattern. Visually, lightcurves created with 
$\lambda \sim 20-30$ look most similar to the observed lightcurve of Cyg X-1. We will 
return to this point later.

\subsection{Wavelet maps for simulated lightcurves}
\label{ss:wasimlc}

We generated 1000 s long lightcurve, $F_N(t)$, with assumed parameters: $\tau=0.067$ s,
$\lambda=20$, $A_0=2500$ cts~s$^{-1}$ and $\nu_2=1.817$ Hz. The mean countrate of our 
simulated time-series yielded to be equal 6081.8 cts~s$^{-1}$ whereas fractional rms 
21.9\%. In the first step we computed PSD to check whether such composed lightcurve is 
able to reproduce observed shape of power spectrum from the PCA data. Result of it has 
been presented in Fig.~\ref{fig:lorentz} by the diamond markers. As one can see, the shape 
and the width of PSD of simulated time-series agrees very well with the main spectral 
component of Cyg X-1 PSD.

Next, from $F_N(t)$ we again chose randomly three sequences, 20 s each, for the wavelet
analysis. The wavelet maps for the simulated lightcurve are shown in 
Fig.~\ref{fig:sim_v6}. 3-D picture of exemplary series is shown in the bottom 
panel of Fig.~\ref{fig:3D}.

The maps are very similar to those obtained from the data analysis. This can 
be even better seen from the quantitative parameters given in Table~\ref{tab1}.
Peaks in the maps for simulated lightcurves also are distributed in a
broad frequency range.

At first glance this result may seem puzzling. Since the basic frequency in the
simulated lightcurve is fixed, we might expect better localization of the
peaks in frequency. However, the random choice of the phase, together with
extremely strong damping, creates an apparent diversity of the shapes among the
consecutive shots, leading to relatively broad peaks, at broad frequency range,
as in the data.
 
Closer inspection suggests also certain systematic differences. Simulated lightcurves 
create slightly fewer peaks in the maps, and the median frequency in the data is somewhat 
higher. This is most probably related to an excess of the power at high frequencies above 
a single Lorentzian peak seen in the data ($\sim$ 8--12 Hz, see Fig.~3). Otherwise, all 
properties of the observed lightcurve are well represented.

We also analyzed simulated lightcurves assuming lower and higher values of
$\lambda$ (with appropriate $A_0$ given by relation (\ref{eq:Alambda})). 
Maps for significantly larger or significantly smaller values of $\lambda$ 
were different from the maps for Cyg X-1. Most noticeably, the activity
level shortened  to 45\% for $\lambda = 5$ and 55\% for $\lambda = 55$;
it was 59.5\% for $\lambda = 20$ and 61.9\% in the data maps. Therefore,
wavelet maps give constraints for $\lambda$, although they are not very 
sensitive to its choice.

We tested the map sensitivity to the adopted bin size in the data and in the simulations. 
Adopting $\Delta t = 2^{-9}$~s (i.e. the minimal bin size available in 
analyzed PCA data set) the noise in the lightcurves (and in the power spectrum) increased 
but the maps were barely changed, as already noticed by Liszka, Pacholczyk \& Stoeger 
(2000b).

\subsection{Maps for lightcurves simulated in the frequency domain}
\label{ss:sim2}

An additional effort was undertaken in order to compare our simulations to the results coming 
from the Timmer \& K\"onig (1997) approach (TK). Namely, their method allows to obtain a 
simulated lightcurve from the PSD of a given shape.

Since a typical PSD of Cyg~X-1 is often well described by a power-law with two breaks
and the three slopes of  
0, 1 and 2, we decided to generate first a TK lightcurve for such a power spectrum. 
The break frequencies were set at 0.1 Hz and 10 Hz, correspondingly, and
again, randomly selected 60 s sequence from 1000 s long time-series ($\Delta t= 0.025$~s) 
was selected for a closer investigation in the 1--10~Hz range. In result we found that
the wavelet maps reveal almost no patterns of variability at the level of significance greater
than 90\%. We estimated the $\kappa$ level for 20 s time-series at only a few per cent.
Only a few peaks (1-3) were able to reach the significance of 99\%.

The same method and approach were used by us to generate the power spectrum assuming the shape of 
the underlying variability due to a Lorentzian damping. We followed the numbers given in Section
\ref{ss:simlc} to perform simulations. As an outcome, surprisingly we found a very good 
agreement of the wavelet peak properties comparing to the results obtained from the wavelet 
maps for the (\ref{eq:lorentz_przepis}) lightcurve. That makes these two techniques essentially
equivalent in the sense of the wavelet map results.

In none of the performed TK simulations we notice any chains of features with a 
systematically increasing frequency along the time axis.

\subsection{Morlet wavelet with non-standard parameters and other wavelet shapes}

Following the most recent analysis of De Moortel, Mundat \& Hood (2004) we tried to 
increase the time resolution at the expense of frequency resolution by adopting smaller 
values of $2\pi\nu_0$. We constructed the Morlet wavelet map both for the data 
(see Fig.~\ref{fig:inne}, top panel) and for the simulated lightcurve (map not shown)
using $2\pi\nu_0=3$.
However, in this case we loose the resolution in frequency. Features become 
more elongated. Again, no pattern in frequency evolution can be seen. The extension of the 
peaks in the time domain decreased. A systematic difference between the observed and the 
simulated lightcurve with respect to the number of localized peaks and the extension of 
active periods remains. Conversely, an attempt in adopting the Morlet parameter 
$2\pi\nu_0>6$ returned in more precise frequency resolution of existing features at the expanse 
of their time localization (map not shown). Also in this case, the statistics for the 
map did not change for the advantage of adopting Morlet function with 
$2\pi\nu_0>6$~\footnote{Please note that Morlet wavelet with $2\pi\nu_0<5$ and $>6$ does not 
meet admissibility condition (Section \ref{sect:wavelets}) and its usage can be justified only for
non-quantitative analysis.}.

We also tried to apply two other wavelet shapes: a Marr and a Paul wavelet. Marr gave a 
very poor resolution in frequency both for the data (Fig.~\ref{fig:inne}, bottom panel) 
and for the simulated lightcurve (map not shown), even worse than Morlet wavelet with 
$2\pi\nu_0=3$. Most of the peaks were so elongated that they extended beyond the studied 
frequency range, making this wavelet rather useless for the studied lightcurves. However, it 
indicated a fast disappearance of the peaks, in time-scales shorter than 0.3~s. Also the 
systematic difference between the observed and the simulated lightcurve was again seen: 
activity periods in Cyg X-1 covered larger fraction of time than in the simulated 
time-series.
    
The Paul wavelet (Fig.~\ref{fig:inne}, middle panel) had effective properties similar to the 
Morlet wavelet with $2\pi\nu_0=3$. Therefore, the use of the canonical Morlet shape with 
$2\pi\nu_0 = 6$ seems to be indeed the most profitable in the context of the X-ray lightcurves.

\subsection{Applicability of a single Lorentzian model to the failed transition state}

Our wavelet analysis generally supports the interpretation of Cyg X-1 variability
during the failed state transition as damped oscillations with fixed underlying frequency.
Additional support comes from the recent paper by Feng, Zhang \& Li (2004). They apply a
different technique to analyze the lightcurve of Cyg~X-1 ({\it w} spectral analysis)
which is sensitive to the shortest time-scales present in the system. The resulting 
time-scales for the hard and the intermediate state are of the same order as the damping 
time-scale of the Lorentzian, consistent with the extension of the detected features on 
our wavelet maps. 

Simulated lightcurves gave maps quantitatively similar to the maps obtained from the
Cyg X-1 lightcurve. The observed PSD contains an excess of power at higher frequencies
(modeled by P03 as an additional Lorentzian peak). Simulated
lightcurves did not contain this additional high frequency input but still reproduced
the maps properly so the dominant Lorentzian is responsible for the character of the
variability both in the frequency domain and in the time--frequency domain.

\subsection{Dominating PSD components in hard, soft and failed soft state}

Our results give a strong support to the interpretation of the PSD through Lorentzian 
peak during the failed state transition. It means that in general case we most probably 
deal with three main {\it physical} components of the PSD with the following properties 
(Nowak 2000; P03):
\vspace*{-5pt}
\begin{itemize}
\item
     Power law component $\nu^{-1}$: this component dominates the PSD at $\nu 
     <10^{-5}$~Hz (Reig, Papadakis \& Kylafis 2002), it extends up to 15--20~Hz when the 
     source is in soft state, it extends into the region $\nu \sim 10^{-3}$--10~Hz but 
     with suppressed amplitude during the transition state and it practically disappears in 
     the hard state;
\item 
     Lorentzian peak $L_1$: present in the hard state but its amplitude strongly decreases 
     during the transition state, accompanied with the rise of the power law in high 
     frequency range;
\item
     Lorentzian peak $L_2$: suppressed only when the the soft state is reached; 
     ratio of the two Lorentzian frequencies is constant ($\sim 7$) although the 
     frequency can change in long time-scale up to a factor of 4. 
\end{itemize}
\vspace*{-5pt}

Results for the coherence function and the Fourier time delays support the view that both 
Lorentzians are generally highly coherent, also between themselves (Nowak et al. 1999) but 
when one of the peaks disappears in the failed state transition, the mean time delay 
between 2--4 keV and 8--13 keV increases to 8 ms (a typical value if measured at 
$\sim 3$~Hz), and the coherence in 3--10 Hz band drops to 0.6 (P03). 
Similar properties were observed by Cui et al. (1997b) (e.g. 30.05.1996) when the 
source was on its way to the completed transition to soft state. 

These components seem to be critical for our understanding of the nature of the observed 
variability although exact fits to the data generally require more components.

\subsection{Physical interpretation of the X-ray variability}
 
\subsubsection{Power Law component $\nu^{-1}$: propagating disk oscillations filtered at 
               the transition radius}
\label{sss:plaw}

There are several independent arguments in favor of the hypothesis that the $\nu^{-1}$
power law component of the PSD is related to the presence of the cold disk. 
This component is exclusively responsible for the observed variability when Cyg~X-1 is in 
the soft state, i.e. radiation spectrum is dominated by the disk emission and we see the 
relativistically broadened iron K$\alpha$ line which gives constraints for the cold disk 
inner radius (e.g. Di Salvo et al. 2001). The time-scales involved cover uniformly a 
large range, including time-scales as long as years which are naturally expected in the 
outer disk. However, there were two puzzles which were not initially understood. First, 
what we see is {\it not the variable disk emission}, since the direct disk emission is 
roughly unchanged but {\it variable Compton component} (Churazov, Gilfanov \& Revnivtsev 
2001; Maccarone \& Coppi 2002). Second, the energy contained in the longest time-scale 
variability is considerable while the gravitational energy at the outer edge of the disk
is small.

The model which successfully explains the situation is the scenario of propagating
perturbations. The idea was nicely outlined by Lyubarskii (1997), although some
basic ideas were already contained in the automaton model of Mineshige et al. (1994).
More sophisticated approach was developed by King et al. (2004). In this model local 
perturbations form in the disk as a result of the action of the  magnetorotational instability 
(MRI; Balbus \& Hawley 1991). Characteristic frequencies of these perturbations at a give 
radius $r$ are somewhat lower than the local Keplerian frequency and can be expressed as:
\be
\nu_{\rm mag} = 20 \left(\frac{5.5}{k_{\rm mag}} \right) 
                \left(\frac{3 R_{\rm Schw}}{r} \right)^{3/2} 
                \left(\frac{20 M_{\odot}}{M}\right) \ \ 
                \mbox{[Hz]} .
\label{eq:mag}
\ee
Most of the authors argued that the inner radius of the disk in the soft state is of 
order of 3 $R_{\rm Schw}$ (e.g. Frontera et al. 2001 give $6^{+4}_{-0} R_{\rm g}$, 
where $R_{\rm g}=0.5 R_{\rm Schw}$) and we adopted this value as a convenient unit. 
The characteristic frequency at this radius reproduces the observed PSD frequency break 
at 20~Hz in the soft state when $k_{\rm mag}=5.5$. The factor of $2\pi$ was included, i.e. 
$\nu = \Omega/(2 \pi)$. Those perturbations propagate inward, to the region where 
dissipation and the production of the hard X-ray emission proceeds. Therefore, the 
accretion rate at the inner dissipation region has a memory of all time-scales present in 
the disk at all radii.

When the source proceeds from a soft state to a hard state, the power law
component is not seen at time-scales shorter than $10^{-5}$--$10^{-3}$~Hz. We
cannot relate this change to a change of the disk inner radius since this would
require a change of the radius by a factor of $10^3$--$10^4$. Analysis of the
data suggest much lower values; e.g. $\sim5$--$35 R_{\rm Schw}$ (di Salvo et al. 2001), 
$10^{+5}_{-4} R_{\rm g}$ (Frontera et al. 2001) or $\sim 100 R_{\rm g}$ (Gilfanov, 
Churazov \& Revnivtsev 2000). However, due to a different treatment of the Compton 
thickness of the inner disk, the values of the radii quoted above should be considered here 
with a caution.

However, a natural explanation of this change is found if we adopt {\it the
strong ADAF principle} after Narayan \& Yi (1995). This is a statement which tells
that {\it whenever an ADAF can form, it does form}. The consequence of this assumption is 
that for a given value of the transition radius, $r_{\rm tr}$, only one value of the 
accretion rate is possible. This fact automatically leads to filtering all time-scales
in the momentary accretion rate shorter than a local viscous time-scale. If
the perturbed accretion rate is somewhat larger than allowed by a current
position of the transition radius, the matter accumulates near $r_{\rm tr}$ while
the accretion rate in the inner flow is unperturbed. Mass accumulation, if
persisting for long enough, leads to a build-up of a cold disk in part of the region
previously occupied by the inner hot flow. The position of the transition radius
moves in, and after this change an inner hot flow can transmit mass at higher rate,
appropriately to the new position of $r_{\rm tr}$. Alternatively, if the perturbed
accretion rate in the disk is somewhat lower than current $r_{\rm tr}$, the inner
hot flow still persists with the same accretion rate at the expense of the
cold disk. Finally, the cold material is used up, the transition radius
moves out to the new position, and the accretion rate in the inner flow decreases
in agreement with new $r_{\rm tr}$. Since both the removal and the buildup of the
cold disk happens in a local viscous timescale all variations in the accretion
rate slower than that are transmitted while faster variations are damped.
The fastest timescale for a transition from a hard to a soft state is also
given by a viscous time-scale, 
\be
\tau_{\rm visc} = 1.5 \times 10^3 
                  \left(\frac{0.01}{\alpha}\right)
                  \left(\frac{0.08 r}{h} \right)^{2} 
                  \left(\frac{r}{50 R_{\rm Schw}}\right)^{3/2} \times \nonumber
\ee
\be
  ~~~~~~~~~~~~~~~~~~~~~~~~~~~~~~~~~~~~~~~~~~~~~~~~~~~~~~~~
  \times \left(\frac{M}{20 M_{\odot}}\right) ~~\mbox{[s]}~
\ee
and it is observationally constrained to be of order of a day, or a fraction of a day, 
consistent with the time-scale $10^{-3}$--$10^{-5}$ Hz. Here we assumed the expected
ratio of the disk thickness to the radius, $h/r$ of order of 0.08 and $\alpha=0.01$ 
after King et al. (2004), for the scaling purposes.

This picture gives a strong observational support to the {\it strong ADAF 
principle}. An independent argument for this kind of behavior was also
found by Czerny, R\' o\. za\' nska \& Kuraszkiewicz (2004) from the study of the 
constraints for an inner radius from Broad Line Region in active galactic nuclei.

During the transition state the accretion rate is most probably much more
strongly enhanced than during the hard state perturbations. Observations show
a sudden decrease in the inner radius. For example, on 1996.05.30, when the state
was yet similar to the failed state transition, the inner radius was estimated
to be $9^{+13}_{-4} R_{\rm Schw}$ (Gierli\' nski et al. 1999).
However, we can suspect that the material piles up rapidly. Instead of
smooth motion of the transition radius clumps of the cold material possibly
enter the inner hot flow region and accrete. Such a direct accretion of
a cold phase may {\it (i)} explain why during the transition, including failed
transition state, the short time-scale perturbations in the accretion flow
start to propagate inward, {\it (ii)} cold moving blobs provide additional
source of soft photons for Comptonization which disrupt the high coherence
seen during the hard state. 

It still remains to explain what kind of instability leads to
the disk destruction when ADAF solution is possible. 
An interesting scenario is discussed by Spruit \& Deufel (2002), but it requires
the initial existence of the hot ions. It may give a hint why we observe a
strong hysteresis effect in some sources but not in Cyg X-1. This hysteresis
(see e.g. Miyamoto et al. 1995; Meyer-Hofmesiter, Liu \& Meyer 2005; 
Zdziarski et al. 2004) leads to a transition from soft to hard state at much lower 
accretion rate than the transition from hard to soft state. In the soft state of
Cyg~X-1 significant fraction of emission is of non-thermal origin. Hot ions are 
always there, and the mechanism of Spruit \& Derfel (2002) may operate while we need 
another mechanism for sources which do not show any hot plasma in the soft state.

\subsubsection{Lorentzian peaks $L_1$ and $L_2$: MRI and dynamical pulsations of the inner ion 
               torus}
\label{sss:l1l2}

The two Lorentzian peaks dominating the hard state are expected to be related to the
inner ion torus. The constant ratio between their frequencies strongly suggest that we
see two kinds of variability coming from the same medium.

Since the inner hot flow must transport the material the same MRI instability is
expected to be operating. Since the inner flow is expected to be geometrically thick,
instability has more global character there, and the dominant frequency is likely
to be determined by the transition radius. This frequency still should be roughly 
reproduced by equation (\ref{eq:mag}):
\be
\nu_{1} = 20 \left(\frac{5.5}{k_{\rm mag}}\right) 
          \left(\frac{3 R_{Schw}}{r_{tr}}\right)^{3/2}
          \left(\frac{20 M_{\odot}}{M}\right)~~\mbox{[Hz]}.
\ee
When the second Lorentzian peak is at 1.7 Hz, the first peak is at 0.25 Hz 
(P03), and such a frequency is reproduced for a reasonable value 
$r_{\rm tr} = 55 R_{\rm Schw}$. In general, the position of the Lorentzian varies. 
Revnivtsev, Gilfanov \& Churazov (2001) showed that the corresponding peak in the power spectrum of 
GX 339-4, described as quasi-periodic oscillation (QPO), strongly correlates
with the other spectral properties of this source, supporting its connection with the change
of the transition radius. 

Higher frequency variations are likely to represent the dynamical pulsation of the inner
hot flow, or traveling sound waves. Global oscillations of this medium, modeled as a 
torus, were studied in a number of papers (e.g. Abramowicz, Calvani \& Nobili 1983; 
Giannios \& Spruit 2004; Montero, Rezzolla \& Yoshida 2004). The typical frequencies of 
{\it p}-modes depend on the details of the model (e.g. assumed radial distribution of the 
angular momentum) but they are roughly of the order of the Keplerian frequency at the 
outer edge of the torus (Rezzolla, Yoshida \& Zanotti 2003). The excitation of the modes 
is due to a radiative interaction of the disk and the torus, while the synchrotron emission 
provides the damping mechanism (Giannios \& Spruit 2004). 

The expected frequency of this mode is roughly:
\be
\nu_{2} = 1.62 \left(\frac{50 R_{\rm Schw}}{r_{\rm tr}}\right)^{3/2} 
          \left(\frac{20 M_{\odot}}{M}\right)~~\mbox{[Hz]}
\ee
so the value of the transition radius $\sim 50 R_{\rm Schw}$ gives the value of the 
frequency roughly corresponding to the observed frequency of the Lorentzian, 1.7 Hz.

During the failed transition state the first Lorentzian peak is strongly damped, and a 
power law components is partially rebuilt, as discussed in Section \ref{sss:plaw}. If the first 
Lorentzian peak is connected with MRI instability, the exchange of power between a power 
law component and the first Lorentzian is natural. Decreased MRI instability in the inner 
flow suppress the accretion of the hot material and the accretion proceeds predominantly 
through cold blobs. It is an interesting question whether the presence of the cold blobs 
prevent MRI instability or the absence of hot material accretion forces the cold phase to 
take over but the eventual compensation of the two mechanisms is quite intuitive. 
Apparently, dynamical pulsations persist as long as the hot inner
torus is still there. If a transition to a soft state is completed, this phase finally 
disappears.

The suggested picture may be too simple since actually a number of instabilities may 
operate in the accretion flow (e.g. Menou, Balbus \& Spruit 2004 and the references 
therein) but we consider our interpretation as a plausible and attractive scenario. 
Clearly, quantitative models would be needed to work out the detailed predictions.

Observationally, an insight may come from the presence of rapid, very energetic events 
reported by Gierli\'nski \& Zdziarski (2003) both during soft and hard state.
It is interesting that a soft state flare no. 13, studied in detail by these authors, was 
found to be well described by damping time-scale (after the peak) of $\tau=(21\pm 6)^{0.7\pm 
0.1}$~ms, i.e. $\tau\simeq 0.07$~s. This damping time-scale is the same as the damping 
time-scale of the $L_2$ Lorentzian, which may indicate that damping is related to magnetic field 
reconnections. Same event observed in the hard state yielded much longer damping 
time-scale, $\sim 0.32$~s which in turn roughly coincides with the damping time-scale of the 
first Lorentzian, $L_1$. Although the exact origin of the flares
is not known, one takes into account sudden conversion of energy accumulated in the disk into
magnetic heating of a hot plasma or fast release of energy due to magnetic field reconnection in 
the flares hung above the disk (Gierli\'nski \& Zdziarski 2003 and references therein).
The difference between the hard state and the soft or failed transition state may be due to
the absence of the cold material in the former case. Therefore, in the hard state the
large scale magnetic loops exist within the torus, while in the soft state or failed 
transition state magnetic field lines are partially frozen into the cold disk or cold blobs.   

\subsection{Wavelet analysis as a tool for tracing accreting matter onto black-holes?}
\label{ss:trace}

It is tempting to associate the observed puzzling chain of peaks in Fig. \ref{fig:5islands}
(mentioned already by us in Section \ref{ss:wamaps}) with a systematic trend.  Closer look
in this figure shows up, in fact, not one but two sequences of peaks roughly of the same slope. 
The first one, most prominent, 
extends between 383--391~s, whereas the second one between 389--397~s. 
Comparison of wavelet map with corresponding PCA lightcurve reveals that all wavelet features 
are associated with some peaks occurring in the same moments in the time-series, as expected.
In addition, a careful inspection of all 20 s long sequences of the PCA observation uncovers 
more similar chains of the same slope. Unfortunately, their number is low i.e. $\sim 14$ per 
whole lightcurve, each one of duration $\le 7$ s. Interestingly, we did not find similar
chains of evolution from lower to higher frequencies in the simulated data.

The time--frequency evolution of some features is not totally unexpected. Both the hot ion
torus material and cold clumps flow onto the black-hole, in the direction of decreasing 
Keplerian time-scales. We can witness a manifestation of the single events like a sequence of 
magnetic field reconnections in an inflowing material. Theoretically, this issue were discussed 
already by Stoeger (1980), however, his considerations were mostly appropriate for events taking 
place below the marginally stable orbit. Here, it seems that observed features 
between 1--10~Hz correspond rather to the range of 69--15~$R_{\rm Schw}$.

If these enigmatic chains of oscillations are related somehow to the processes of 
X-ray emission coming from the material accreting onto black-holes, then, in our opinion, a 
wavelet analysis may become a promising tool in their spotting. However, any judgement on
the reality and physical origin of the observed {\it trends} can be only done after
much more quantitative analysis. The number of the detected {\it chains} must be large and
statistically significant.


\section{Conclusions}

For the first time, we applied the wavelet multi-resolution approach in order to perform 
the complex 
studies of aperiodic X-ray variability of Cyg~X-1. Considered data set corresponded to its 
failed state transition on 1999.12.05. On that day, the Fourier power spectrum revealed a 
broad-band peak between $\sim$1--10 Hz, well described by a single Lorentz function of unclear 
origin. On the contrary to the Fourier method, the wavelet analysis provided an 
excellent inspection of 
the frequency content of the signal as well as allowed for simultaneous spotting the 
time evolution of the extracted features.

Our results can be summarized as follows: 
\begin{itemize}
\vspace*{-4pt}
\item Cyg X-1 lightcurve analyzed in the time--frequency plane contained randomly appearing 
flare-like structures of duration  $\sim 0.5 $~s and of the average median of the peak 
frequency width $\Delta\log\nu_p\simeq 0.14$; 
\item observed Lorentzian shape forms through summation of these time-localized structures;
\item most of the Cyg X-1 lightcurve properties are well reproduced by the simulated
lightcurve composed of a series of randomly occurring flares. In our model, each flare was 
described as a damped oscillator with variable amplitude and phase but fixed excitation 
frequency and time-scale of decay;
\item Cyg X-1 wavelet maps may show insignia of an inflow while the maps representing the 
simulated lightcurve do not seem to show similar tendency. However, no statistical significance 
was assigned to this hypothesis.
\end{itemize}


\section*{ACKNOWLEDGMENTS}

The work dedicated in memory of Prof. Edward Wilk. We thank Chris Torrence for many helpful
and friendly discussions on wavelets, and the referee and Piotr \.Zycki for valuable 
remarks. The wavelet software was provided by Christopher Torrence and Gilbert P. Compo 
and it is available at http://paos.colorado.edu/research/wavelets. Part of this work was 
supported by grants 2P03D00322 and PBZ-KBN-054/P03/2001 of the Polish
State Committee for Scientific Research. We also acknowledge the use 
of data obtained through the HEASARC online service provided by NASA/GSFC.



\appendix

\section{Discrete form of CWT in terms of Fourier transform}
\label{s:dcwt}

Let us express a continuous wavelet transform as the convolution of the Fourier 
transforms of the lightcurve and wavelet function in the following way:
\be
   w_{a,b} = \int_{-\infty}^\infty \hat{x}(\nu) \hat{\psi}^\star_{a,b}(\nu) d\nu .
\ee
The Fourier transform of $\psi_{a,b}$ is given by:
\be
   \hat{\psi}_{a,b}(\nu) = \frac{1}{\sqrt{a}} \int_{-\infty}^\infty \psi
                                 \left(\frac{t-b}{a}\right) e^{-i2\pi\nu t} dt .
\label{a:hatpsi}
\ee
Substituting $(t-b)/a$ by $\eta$ equation (\ref{a:hatpsi}) evaluates into:
\be
   \hat{\psi}_{a,b}(\nu) = \sqrt{a} e^{-i2\pi\nu b}
   \int_{-\infty}^\infty \psi(\eta) e^{-i2\pi a \nu \eta} d\eta
\label{a:hatpsi2}
\ee
where a constant component was separated out. As one can see, an integral expression can be 
considered as the Fourier transform of a wavelet function where $2\pi a\nu$ is a rescaled 
angular frequency. Thus it is possible to denote (\ref{a:hatpsi2}) shortly:
\be
   \hat{\psi}_{a,b}(\nu) = \sqrt{2\pi a}\, e^{-i2\pi\nu b} \hat{\psi}(2\pi a\nu)
\ee
and therefore a CWT as the inverse Fourier transform of the product:
\be
   w_{a,b} = \sqrt{2\pi a} \int_{-\infty}^\infty \hat{x}(\nu) \hat{\psi}^\star(2\pi a\nu) 
   e^{i2\pi\nu b} d\nu
\ee
where complex conjugate of the Fourier transform of $\psi$ function was taken into account.

A discrete Fourier transform of $x_k$ is given by 
\be
   \hat{x}_j = N_{\rm obs}^{-1} \sum_{k=0}^{N_{\rm obs}-1}
               (x_k - \bar{x}) e^{-i2\pi j k/N_{\rm obs}}
\ee
where $j$ denotes a 
frequency index. Performing a discrete notation of the time-shifting factor as $b=k\Delta t$ 
where $\Delta t$ stands for the sampling time of a lightcurve, one can write down a {\it 
discrete form of continuous wavelet transform in terms of the Fourier transform} as:
\be
   w_k(a) = \left(\frac{2\pi a}{\Delta t} \right)^{1/2} \sum_{j=1}^{N_{\rm obs}}
          \hat{x}_j \hat{\psi}^\star(2\pi a\nu_j) e^{i2\pi\nu_j k\Delta t}
\label{a:wka}
\ee
where frequency $\nu_j$ equals:
\be
\nu_j =
\begin{cases}
\ \ \,  j/(N_{\rm obs}\Delta t) & \text{for $j\leq N_{\rm obs}/2$} \\
  -j/(N_{\rm obs}\Delta t) & \text{for $j> N_{\rm obs}/2$}
\end{cases}
\ee
and the factor $(2\pi a/\Delta t)^{1/2}$ ensures that 
wavelet will keep the same energy at every scale $a$. Because of the finite duration of the 
signal, the proper choice of scales must be done. The largest scale would correspond to 
the lightcurve span of $T$ whereas the smallest one to ought to be an equivalent of the 
Fourier Nyquist frequency, i.e. $a_0=2\Delta t$. In fact, a selection of scales in this 
region can be chosen arbitrarily with a step in scale not smaller than $a_0$. However, in 
case of very fine signal when a few decades of scales are wished to be covered, one can 
build the wavelet transform choosing:
\be
    a_m = a_0 2^{m\Delta m} \ \ \ \ m=0,...,M
\ee
where
\be
    M=\Delta m^{-1} \log_2(N_{\rm obs}\Delta ta_0^{-1})
\ee
(TC98) and thus rewriting (\ref{a:wka}) in the following way:
\be
   w_k(a_m) = \left(\frac{2\pi a_m}{\Delta t} \right)^{1/2} \sum_{j=1}^{N_{\rm obs}}
          \hat{x}_j \hat{\psi}^\star(2\pi a_m\nu_j) e^{i2\pi\nu_j k\Delta t} .
\label{a:wkam}
\ee

\bsp
\label{lastpage}
\end{document}